\documentclass[12pt,a4paper]{article} 
\usepackage{epsfig}

\newcommand{\<}{\langle}  
\renewcommand{\>}{\rangle}

\newcommand{\beq}{\begin{equation}}   
\newcommand{\eeq}{\end{equation}}   
\newcommand{\beqn}{\begin{eqnarray}}  
\newcommand{\eeqn}{\end{eqnarray}}   
\newcommand{\nn}{\nonumber}

\newcommand{\half}{\frac{1}{2}}   
   
\usepackage{chir_layout}   

\def\rmii{a}
\def\liv{b}
\def\hum{c}

\begin{document}   

\begin{titlepage}
{
 \vspace{-3.5cm}
\normalsize
\hfill \parbox{85mm}{ROM2F/2009/12, LTH834, HU-EP-09/32}}\\[10mm]
\begin{center}
  \begin{Large}
    \textbf{O($a^2$) cutoff effects in  lattice  \\ Wilson fermion simulations
            \unboldmath} \\
  \end{Large}
\end{center}

\begin{figure}[!h]
  \begin{center}
    \includegraphics[scale=1.0]{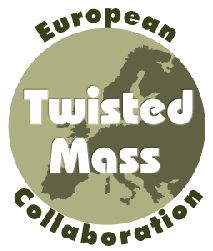}
  \end{center}
\end{figure}

\baselineskip 20pt plus 2pt minus 2pt
\begin{center}
  \textbf{
    P.~Dimopoulos$^{(\rmii)}$,
    R.~Frezzotti$^{(\rmii)}$,
    C.~Michael$^{(\liv)}$,\\
    G.C.~Rossi$^{(\rmii)}$,
    C.~Urbach$^{(\hum)}$}\\
\end{center}

\begin{center}
  \begin{footnotesize}
    \noindent

$^{(\rmii)}$ Dip. di Fisica, Universit{\`a} di Roma Tor Vergata and INFN,\\
 Sez. di Roma Tor Vergata, Via della Ricerca Scientifica, \\ I-00133 Roma, Italy
\vspace{0.2cm}

$^{(\liv)}$ Theoretical Physics Division, Dept. of Mathematical Sciences,
\\University of Liverpool, Liverpool L69 7ZL, UK
\vspace{0.2cm}

$^{(\hum)}$ Institut f\"ur Elementarteilchenphysik, Fachbereich Physik,
\\ Humbolt Universit\"at zu Berlin, D-12489, Berlin, Germany
\vspace{0.2cm}

\vspace{0.2cm}

  \end{footnotesize}
\end{center}

\begin{abstract}{
In this paper we propose to interpret the large discretization artifacts affecting the
neutral pion mass in maximally twisted lattice QCD simulations as O($a^2$) effects
whose magnitude is roughly proportional to the modulus square of the (continuum)
matrix element of the pseudoscalar density operator between vacuum and one-pion
state. The numerical size of this quantity is determined by the dynamical mechanism
of spontaneous chiral symmetry breaking and turns out to be substantially
larger than its natural magnitude set by the value of $\Lambda_{\rm QCD}$.}
\end{abstract}

\end{titlepage}
\vfill
\newpage  

\section{Introduction}  
\label{sec:INTRO}   

In recent simulations of maximally twisted lattice QCD (Mtm-LQCD)~\cite{TM}
with $N_f=2$ dynamical light quarks, numerical results~\cite{LET,PROC} on the 
neutral pseudoscalar meson mass at different lattice spacings (in the range from 
0.10 to 0.065~fm) and quark masses (corresponding to charged pseudoscalar meson 
masses ranging from 600 to 300~MeV) show large cutoff effects.
The latter appear at fixed lattice spacing in the form of sizeable
deviations from the expected continuum QCD chiral behaviour.  

This finding is in striking contrast with the smallness of cutoff effects observed not only 
in the mass of the charged pions (which is related through the Goldstone theorem to exactly 
conserved lattice currents~\cite{FMPR}), but also in all the other hadronic observables 
so far measured by the European Twisted Mass (ETM) Collaboration (see sect.~\ref{sec:EXDA} 
and refs.~\cite{PROC,ETMCLQM,BARY,ETMC_long,TREE,PFF,FD}). Quite remarkably small 
lattice artifacts are found even in matrix elements where the neutral pion is involved (see 
Table~\ref{tab:results} below). 

In Mtm-LQCD the additive cutoff effect in the squared mass of the neutral pion 
is expected on general grounds~\cite{FR1} to be of the order $a^2\Lambda_{\rm QCD}^4$,
whereas the leading additive corrections to the squared mass of the charged pion
are~\cite{SCOR,SHWU,FMPR} only of order $a^4\Lambda_{\rm QCD}^6$ and 
$a^2\mu_q\Lambda_{\rm QCD}^3$ ($\mu_q$ denotes the bare quark mass). 
The neutral vs.\ charged pion (squared) mass splitting, 
$\Delta m_\pi^2|_{L}^{\rm Mtm}=m_{\pi^3}^2|_{L}-m_{\pi^\pm}^2|_{L}$,
in the small quark mass region is then dominated by O($a^2\Lambda_{\rm QCD}^4$) 
terms. Numerical data (see Fig.~\ref{fig:figMPISPLIT} and Table~\ref{tab:results}) at two 
different lattice spacings and for charged pion masses below 500~MeV are indeed 
consistent with the expected behaviour of $\Delta m_\pi^2|_{L}^{\rm Mtm}$, 
but with a large coefficient in front of the O($a^2\Lambda_{\rm QCD}^4$) 
correction~\footnote{The issue of the $N_f$ dependence of this coefficient is beyond 
the scope of this paper and will be discussed elsewhere.}. 
For instance, taking $\Lambda_{\rm QCD}=250$~MeV, one finds 
$\Delta m_\pi^2|_{L}^{\rm Mtm}\sim -50 a^2\Lambda_{\rm QCD}^4$.

On a more general ground, the question of the numerical size of the neutral vs.\ charged pion 
mass splitting is an important issue to assess the viability of the Mtm-LQCD approach. This 
is so mainly because of the established relation~\cite{SCOR,META} between the magnitude
of the pion mass splitting and the strength of the metastabilities detected in the theory at 
coarse lattice spacings~\cite{META_NUM,XLF1,XLF2,XLFQUEFLB}. 

One may suspect that the size of $\Delta m_\pi^2|_{L}^{\rm Mtm}$ represents the 
generic magnitude of the isospin breaking effects inherent in the twisted form of the action.  
This is not so, however. Numerical data in several other observables indicate, in fact, that 
in general isospin breaking artifacts are pretty small, with a relative magnitude in line 
with their naive order of magnitude estimate. 

The main purpose of this paper is to provide an explanation for this peculiar pattern 
of isospin breaking effects, and in particular for the fact that only the neutral vs.\ charged 
lattice pion mass splitting among all the other O($a^2$) effects is large. 

Relying on arguments based on the Symanzik analysis~\cite{SYM,LSSW} of lattice artifacts, 
we are able to identify the form of the leading O($a^2$) corrections that in Mtm-LQCD 
affect the value of the squared mass of the neutral pseudoscalar meson. Interestingly 
one can give an approximate evaluation of these effects finding that they are proportional 
to the modulus square of the continuum matrix element of the isotriplet pseudoscalar 
density operator between vacuum and one-pion state. The latter is a dynamically large 
number determined by the mechanism of spontaneous chiral symmetry breaking. 

Lattice artifacts of course also depend on the many unknown coefficients multiplying the 
operators occurring in the Symanzik expansion. However, the numerical evidence provided 
in this paper (see sect.~\ref{sec:EXDA}) and in refs.~\cite{PROC,BARY} about the fact 
that all the other so far measured observables exhibit only small lattice artifacts should be 
taken as a strong indication that the coefficients multiplying the (quark-mass independent) 
operators of the Symanzik effective Lagrangian of relevance here (operators of dimension 
5 and 6) are not unnaturally large, at least in the gauge coupling regime explored up to now 
in ETMC simulations. Indeed, were not this the case, one would have seen large cutoff effects 
in some other physical observables besides the neutral pion mass.

A preliminary version of this investigation was presented some time ago in~\cite{REGFR}. 
Ideas and a few results along the line of reasoning developed in this paper have been already 
put forward in~\cite{MUSC,SCOR,SHWU} and in~\cite{RING}.

\subsection{Plan of the paper}
\label{POP} 

The plan of the paper is as follows. In sect.~\ref{sec:PMMTM} we illustrate the 
nature and the structure of O($a^2$) artifacts in Mtm-LQCD and how they affect 
charged and neutral pion masses. In sect.~\ref{sec:EXDA} we give numerical evidence 
that among the many observables recently measured by the ETM Collaboration only 
the pion mass splitting seems to display large cutoff effects. Conclusions can be found 
in sect.~\ref{sec:CONC}. More technical issues are discussed in Appendices. In Appendix~A  
we review the notion and the properties of the Symanzik expansion for the 
description of the cutoff effects of LQCD with Wilson fermions (either maximally twisted 
or untwisted) and how, depending on the value of the twisting angle, its form is affected 
by the way the critical mass is determined. In Appendix~B we give the structure of 
the Symanzik expansion of the two-point pseudoscalar correlator from which the 
formula for the O($a^2$) discretization errors affecting the pion masses can be derived.  
Finally in the (long) Appendix~C we give details on the theoretical analysis and numerical 
estimate of the matrix elements controlling the magnitude 
of certain cutoff corrections affecting  
neutral and charged pion masses.

\section{Neutral and charged pion mass in Mtm-LQCD}
\label{sec:PMMTM}

In this section we want to describe and compare the structure of the lattice artifacts affecting 
the neutral and the charged pion mass in Mtm-LQCD. The key formulae yielding the O($a^2$) 
corrections to the charged and neutral squared pion mass are immediately derived by writing 
down the leading correction to the pion (rest) energy 
induced by the O($a^2$) terms of the Symanzik expansion. In Mtm-LQCD one gets ($b=\pm, 3$)
\beqn
\hspace{-0.4cm}&&m^2_{\pi^b}\Big{|}_L=
m^2_{\pi}+a^2\Big{[}\langle\pi^b(\vec 0)|{\cal L}_6^{\rm Mtm}(0)|\pi^b(\vec 0)\rangle\Big{|}_{\rm cont}+\nn\\
\hspace{-0.4cm}&&-\frac{1}{2}\langle\pi^b(\vec 0)|\int d^4x\,{\cal L}_5^{\rm Mtm}(x)
{\cal L}_5^{\rm Mtm}(0)|\pi^b(\vec 0)\rangle\Big{|}_{\rm cont}\Big{]}+{\rm O}(a^2m_{\pi}^2,a^4)\, ,
\label{MP3}
\eeqn
where the form of ${\cal L}_5^{\rm Mtm}$ and ${\cal L}_6^{\rm Mtm}$ is detailed in Appendix~A.
Eq.~(\ref{MP3}) complies with elementary lowest order perturbation theory formulae and Lorentz 
covariance. For completeness in Appendix~B we give another derivation of eq.~(\ref{MP3}) starting from 
the Symanzik expansion of the Fourier Transform of the two-point pseudoscalar correlator (no sum over $b$), 
\beq
\Gamma_L^b(p)=a^4\sum_x e^{ipx}\langle P^b(x)P^b(0)\rangle\Big{|}_L\, ,
\label{P3P3T}
\eeq
where $P^b=\bar\psi i\gamma_5 (\tau^b/2) \psi$ and $\tau^b$, $b=1,2,3$, are the Pauli matrices. 

The main conclusions of this paper are derived from the following two observations concerning 
the structure of eqs.~(\ref{MP3}).

1) The pion matrix elements $\langle\pi^b(\vec 0)|{\cal L}_6^{\rm Mtm}(0)|
\pi^b(\vec 0)\rangle|_{\rm cont}$ are O(1) for $b=3$, but only O($m_\pi^2$) for $b=\pm$
(see Appendix~B), viz.
\beqn
&&\langle\pi^\pm(\vec 0)|{\cal L}_6^{\rm Mtm}(0)|\pi^\pm(\vec 0)\rangle\Big{|}_{\rm cont}={\rm{O}}(m_\pi^2)\, ,\label{DEFPM}\\
&&\zeta_\pi\equiv\langle\pi^3(\vec 0)|{\cal L}_6^{\rm Mtm}(0)|\pi^3(\vec 0)\rangle\Big{|}_{\rm cont}={\rm{O}}(1)\, .\label{DEF1}
\eeqn

2) Upon use of the optimal critical mass~\cite{FMPR}, one can show (see Appendix~C) that the matrix element 
\beq 
\Delta_{55}^b=-\frac{1}{2}\langle\pi^b(\vec 0)|\int d^4x\,{\cal L}_5^{\rm Mtm}(x)
{\cal L}_5^{\rm Mtm}(0)|\pi^b(\vec 0)\rangle\Big{|}_{\rm cont}
\label{D55}
\eeq
is parametrically of O($m_\pi^2$) in the chiral limit for $b=\pm$, and numerically negligible 
(in modulus) compared to $|\zeta_\pi|$ in the case of $b=3$. 

The direct consequence of these statements is that close to the chiral limit there are no 
additive O($a^2$) terms affecting the value of the lattice charged squared pion mass. The 
neutral squared pion mass instead receives non-vanishing corrections at this order. In formulae one gets
\beqn
\hspace{-1.5cm}&&m^2_{\pi^\pm}\Big{|}_L= m^2_{\pi}+
a^2\Delta_{55}^{\pm}+ {\rm O}(a^2m_{\pi}^2,a^4)=
m^2_{\pi}+ {\rm O}(a^2m_{\pi}^2,a^4)\, ,\label{DMPPM}\\
\hspace{-1.5cm}&& m^2_{\pi^3}\Big{|}_L=m^2_{\pi}+
a^2\big{[}\zeta_\pi+\Delta_{55}^{3}\big{]}+ {\rm O}(a^2m_{\pi}^2,a^4)
\, .\label{DMP3}
\eeqn
One can also show that in $\Delta_{55}^b$ the numerically dominant O($m_\pi^2$) terms, 
coming from the insertion of one-pion states in~(\ref{D55}), are actually independent of the isospin 
index $b$. As a result what is left in the difference $\Delta_{55}^{3}-\Delta_{55}^{\pm}$ is a tiny 
correction, so that for the pion squared mass difference we can finally write down the simple formula 
\beq
\Delta m_\pi^2\Big{|}_{L}^{\rm Mtm}=m_{\pi^3}^2\Big{|}_{L}-m_{\pi^\pm}^2\Big{|}_{L}
\simeq a^2\zeta_\pi+{\rm O}(a^2m_{\pi}^2,a^4)\, ,
\label{DM}
\eeq
where the symbol $\simeq$ is to remind that the (negligibly small) $\Delta_{55}^{3}-\Delta_{55}^{\pm}$ 
correction has been dropped. 

In the rest of the paper we want to give a proof of the propositions 1) and 2) above and provide 
a numerical estimate of $\Delta_{55}^{3,\pm}$ and $\zeta_\pi$. In this section we shall see 
in particular that for dynamical reasons $\zeta_\pi$ is large compared to its natural value 
($\sim \Lambda_{\rm QCD}^4$), providing in this way an explanation for the fairly big value of 
$\Delta m^2_{\pi}|_L^{\rm Mtm}$ measured in numerical simulations~\cite{LET,PROC}.

The arguments leading to the eqs.~(\ref{DMPPM})--(\ref{DMP3}) and 
the statements above are quite elaborated. To keep the line of reasoning as 
straight as possible, we have deferred them to Appendix~B (derivation of 
eqs.~(\ref{DMPPM}) and~(\ref{DMP3})) and Appendix~C (expression and 
estimate of $\Delta_{55}^{3,\pm}$). 

\subsection{Estimating O($a^2$) lattice artifacts in $\Delta m^2_{\pi}|_L$} 
\label{sec:ELAMP3}

Relying on the results of Appendix~C about the numerical irrelevance 
of $\Delta_{55}^b$ (either because parametrically of O($m_\pi^2$), 
for $b=\pm$, or because numerically small, for $b=3$), we only need to evaluate 
$\zeta_\pi$ in (the chiral limit of) continuum QCD in order to estimate the size  
of the O($a^2$) artifacts in~(\ref{DMP3}) and~(\ref{DM}). This can be done 
under the assumption that a sufficiently accurate order of magnitude estimate of this hadronic parameter 
can be obtained in the vacuum saturation approximation (VSA). Quenched LQCD studies give support to the 
assumption that VSA works well for matrix elements of four-fermion operators between pseudoscalar 
states~\cite{DON,LELL06}. It is very likely that this remains true also in the unquenched theory.
Indications in this sense are actually born out by recent studies with two or three dynamical 
flavours~\cite{RBC,BK}. 

\subsubsection{The theoretical argument}
\label{sec:TA}

The evaluation of $\zeta_\pi$ (see eq.~(\ref{DEF1}))
requires the estimate of the matrix elements of the operators in 
eq.~(\ref{L6gen}) (rotated from the twisted to the physical quark basis) between zero 
three-momentum neutral pion states. This is done in two steps. 

I) By the use of classical soft pion theorem's (SPT's)~\cite{CCW,WEIN}
one recognizes that some operators have O(1) neutral pion matrix elements, while 
others vanish proportionally to $m_\pi^2$. The latter are those that are invariant 
under axial-$\tau^3$ transformations. The former are
\beqn
\begin{array}{ll}
\frac{1}{4} (\bar\psi i \gamma_5\tau^3\psi)\,(\bar\psi i \gamma_5\tau^3\psi)\, ,
& \qquad \frac{1}{4} (\bar\psi\psi)\,(\bar\psi\psi)\, ,
\label{O1}\\
\frac{1}{4} (\bar\psi\tau^3\psi)\,(\bar\psi\tau^3\psi)\, ,
& \qquad \frac{1}{4} (\bar\psi i \gamma_5\psi)\,(\bar\psi i \gamma_5\psi)\, 
,\label{O2}\\
\frac{1}{4} (\bar\psi\sigma_{\mu\nu}\tau^3\psi)\,(\bar\psi\sigma_{\mu\nu}\tau^3\psi)\, ,
& \qquad \frac{1}{4} (\bar\psi\sigma_{\mu\nu} i\gamma_5\psi)\,(\bar\psi\sigma_{\mu\nu} i\gamma_5\psi)\, 
,\label{O3}\\ 
\frac{1}{4} (\bar\psi\gamma_\mu\gamma_5\tau^1\psi)\,(\bar\psi\gamma_\mu\gamma_5\tau^1\psi)\, , 
& \qquad \frac{1}{4} (\bar\psi\gamma_\mu\tau^2\psi)\,(\bar\psi\gamma_\mu\tau^2\psi)\, 
,\label{O4}\\ 
\frac{1}{4} (\bar\psi\gamma_\mu\gamma_5\tau^2\psi)\,(\bar\psi\gamma_\mu\gamma_5\tau^2\psi)\, ,
& \qquad \frac{1}{4} (\bar\psi\gamma_\mu\tau^1\psi)\,(\bar\psi\gamma_\mu\tau^1\psi)\, .
\label{O5}
\end{array}
\eeqn 
The operators~(\ref{O5}) are obtained from those of eq.~(\ref{4FinL6}) after
use of the chiral rotation~(\ref{CTPB}). We note that SPT's relate through the trivial numerical factor $-1$ 
the matrix elements between neutral single-pion states of the two operators in each 
line~\footnote{For the reader convenience we recall that SPT's amount to the relation 
$if_\pi\langle \alpha+\pi^b(\vec 0)|O^c|\beta\rangle=\langle \alpha|[Q_A^b,O^c]|\beta\rangle$,
where $Q_A^b$ is the axial charge with isospin index $b$. External states must carry isospin 
indices such that the two matrix elements are not trivially vanishing. We also note the crossing symmetry relation 
$\langle\alpha+\pi^b(\vec 0)|O^c|\beta\rangle=\langle\alpha|O^c|\beta +\pi^b(\vec 0)\rangle $. 
For an introduction to SPT's and a more detailed discussion (including limitations in their use
if intermediate states with the quantum numbers of $O^c | \alpha \rangle$, or $O^c | \beta \rangle$
happen to be degenerate in energy with the states $| \alpha \rangle$ or $| \beta \rangle$) 
see e.g.~\cite{DGH}, sect.~IV-5, and references therein.}. 
Not surprisingly this remark implies that $\zeta_\pi$ would vanish if we were to deal with
an exactly chiral invariant lattice formulation. In fact, such a formulation would be described
by a Symanzik effective Lagrangian where the two operators in each line of the list~(\ref{O5}), 
being related by a chiral transformation, would necessarily appear with identical coefficients. 

II) In VSA one can give an estimate of the value of $\zeta_\pi$ (eq.~(\ref{DEF1})) as 
follows. First of all, one has to rewrite the matrix elements of the operators~(\ref{O5}) between
zero-momentum neutral pion states in the form that is obtained reducing 
external pions by the use of SPT's~\footnote{In doing so we are enforcing consistency 
with SPT's in using VSA-based estimates of the matrix elements of the operators~(\ref{O5}).
This step is necessary because it is not guaranteed that VSA estimates comply with SPT relations.}.
In this way one checks that only the pion matrix elements of the two operators
in the first line of the list~(\ref{O5}) are not vanishing in VSA. Their actual contribution 
to $\zeta_\pi$ depends of course on the magnitude of the Symanzik coefficients 
(dimensionless functions of the bare gauge coupling) by which the two operators are
multiplied in the expression of ${\cal L}_6^{\rm Mtm}$. In the gauge coupling 
regime of interest for large volume simulations of Mtm-LQCD,  these coefficients 
are expected to be O(1) quantities, but their value (and sign) is otherwise unknown. 
We must thus limit ourselves to an order of magnitude estimate of $|\zeta_\pi|$ for which, 
relying on the line of reasoning outlined above, we write 
\beq
|\zeta_\pi|\stackrel{{\rm VSA}}{\sim}  \frac{2 \hat \Sigma^2}{f_\pi^2} 
\simeq 2 \hat G_\pi^2 \, ,
\label{OAQT}\eeq
where $f_\pi \simeq 92.4$~MeV. The quantities $\hat \Sigma$ and $\hat G_\pi$ denote the continuum 
renormalized chiral condensate and the vacuum-to-pion matrix element of the pseudoscalar density, with 
\beqn
&& G_\pi = \<\Omega|P^b|\pi^b({\vec 0})\>\Big{|}_{\rm cont} \, , 
\qquad P^b= \bar\psi \frac{i}{2} \gamma_5 \tau^b \psi  \, , \label{G-SIGMA-DEF}\\
&&\Sigma = \<\Omega| \frac{1}{2} S^0 | \Omega\>\Big{|}_{\rm cont} \, , 
\qquad \quad \,\, S^0 = \bar\psi \psi \, . \label{P-S-DEF}
\eeqn
The first relation, as explicitly indicated, comes from a direct use of SPT's and vacuum insertion. 
The second holds up to O($m_\pi^2$) corrections and comes from combining the 
Gell-Mann--Oakes--Renner formula~\cite{GMOR}
\beq
2\hat \mu_q \hat\Sigma = f_\pi^2 m_\pi^2 + {\rm O}(m_\pi^4)
\label{GMORF}
\eeq
with the continuum Ward--Takahashi identity (WTI)
\beq
2\hat\mu_q \hat{G}_\pi = f_\pi m_\pi^2\, .
\label{CWTI}
\eeq
We conclude this subsection by stressing that sign of $\zeta_\pi$, which, as we said,
depends on the operator coefficients in ${\cal L}_6^{\rm Mtm}$, is not predicted by
our arguments and is to be learned from numerical simulation experience.

\subsubsection{Numerics for eq.~(\ref{OAQT})}
\label{sec:NU}

A numerical evaluation of $\hat{G}_\pi$, or equivalently of $\hat \Sigma / f_\pi$,  
can be obtained in various ways, i.e.\ either exploiting the direct lattice measurement  
of $\hat{G}_\pi$ or by using (at the physical pion point) the continuum WTI~(\ref{CWTI}). 

I) The direct lattice measurements of $\hat{G}_\pi$  carried out in 
ref.~\cite{LET,PROC,PFF} at different lattice resolutions give
in the continuum and chiral limit 
\beq
\hat G_\pi(\overline{MS},\,2\,{\rm GeV}) \sim (490 \; {\rm MeV})^2 \, , \quad
[-\hat \Sigma(\overline{MS},\,2\,{\rm GeV})]^{1/3} \sim 275 \; {\rm MeV} \!\!\label{EST}
\eeq
with a total error on $\hat G_\pi$ not larger than 5\%. In order to arrive at these numbers the 
renormalization constant of the operator $P^b$, computed in the RI-MOM scheme~\cite{RIMOM} 
and converted to the $\overline{MS}$ scheme, was employed~\cite{Dim_etal_Z}.

II) A second, independent determination of $\hat{G}_\pi$ can be obtained by making 
reference to only continuum quantities if eq.~(\ref{CWTI}) and the PDG~\cite{PDG08} estimate 
$\hat \mu_q(\overline{MS},\,2\,{\rm GeV})=[(\hat{m}_u+\hat{m}_d)/2](\overline{MS},\,2\,{\rm GeV})\sim 3.8$~MeV 
are used. In this way one gets at the physical pion mass 
\beq
\hat G_\pi(\overline{MS},\,2\,{\rm GeV}) \sim (470 \; {\rm MeV})^2 \, .
\label{HATG}
\eeq
Taking into account the small error in the extrapolation from the chiral point to the physical 
pion and the uncertainty on the light quark mass value, we see that the two estimates of $\hat G_\pi$ 
are in very good agreement with each other.

The conclusion of this analysis is that numerically the estimate~(\ref{DM}) of the neutral to charged 
squared mass shift is substantially larger than its natural size, $a^2\Lambda_{\rm QCD}^4$, 
by a factor (see eqs.~(\ref{OAQT}), (\ref{EST}) and~(\ref{HATG}))  
of the order of $2 \hat{G}_\pi^2/\Lambda_{\rm QCD}^4 \sim 25 \div 30$.

\subsection{Numerical results for $\Delta m^2_{\pi}|_L$} 
\label{sec:MPISPLIT}

In this section we wish to summarize the simulation results for $\Delta m^2_{\pi}|_L$ 
that have been obtained from $N_f=2$ ETMC ensembles at two different lattice resolutions,
$a^{-1} \simeq 2.3$~GeV and $a^{-1} \simeq 2.8$~GeV (corresponding to $\beta=3.9$ and
$\beta=4.05$, respectively (see sect.~\ref{sec:NUM2} and Table~\ref{tabC} for details), 
and compare them with the number expected on the basis of eq.~(\ref{OAQT}) for $\zeta_\pi$ 
as well as with the estimate of $\Delta_{55}^3-\Delta_{55}^\pm$ 
derived in Appendix~C  (see eq.~(\ref{DDT})).


\begin{figure}[!hbt]
\centerline{\includegraphics[scale=0.50,angle=-90]{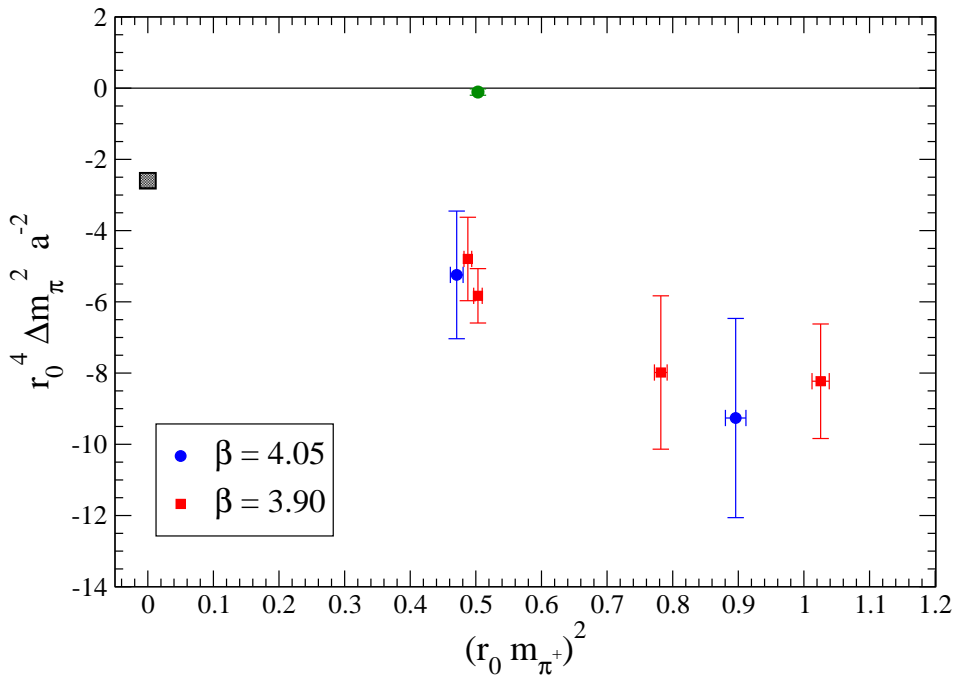}}
\caption{Data for $(r_0/a)^2 r_0^2 \Delta m^2_{\pi}|_L$ vs.\ $r_0^2 m_{\pi^\pm}^2|_L$ 
are compared with the estimate of $r_0^4 \zeta_\pi $ (shaded square)
and $r_0^4 (\Delta_{55}^3-\Delta_{55}^{\pm})$ (green dot).}
\label{fig:figMPISPLIT}
\end{figure}

In Fig.~\ref{fig:figMPISPLIT} we display the lattice results 
for $(r_0/a)^2 r_0^2 \Delta m^2_{\pi}|_L $ 
vs.\ $r_0^2 m_{\pi^\pm}^2|_L$ for the six gauge configuration ensembles specified in
Table~\ref{tabC}. The factor of $(r_0/a)^2$ in the vertical axis has been inserted~\footnote{According 
to ref.~\cite{PROC}, we use $r_0/a=5.22(2)$ at $\beta=3.9$ and $r_0/a=6.61(3)$ at $\beta=4.05$.}
to be able to combine together data from $\beta=3.9$ and $\beta=4.05$, thus giving  
an impression of the quality of the scaling behaviour of $\Delta m^2_{\pi}|_L$.  
In the same figure we also report the value of $r_0^4 \zeta_\pi$ (shaded square), 
which from eqs.~(\ref{OAQT}), (\ref{EST}), (\ref{HATG}) and $r_0 \simeq 0.44$~fm 
is computed to be~\footnote{The sign of $\zeta_\pi$, which appears in eq.~(\ref{zetapi_th}) 
has been inferred from the numerical evidence that $m_{\pi^3} < m_{\pi^\pm}$.}
\beq
r_0^4 \zeta_\pi  \sim - (2.4 \div 2.8) \quad  {\rm @} \; \mu_q = 0 \, .
\label{zetapi_th}
\eeq
The uncertainty in~(\ref{zetapi_th}), which is dominated by the 
approximations in the theoretical arguments given in sect.~\ref{sec:TA}, 
corresponds to the size of the symbol used in the figure. Finally the green dot in 
Fig.~\ref{fig:figMPISPLIT} represents the result of the estimate~(\ref{DDT}) 
we give of $\Delta_{55}^3-\Delta_{55}^\pm$ expressed in $r_0$-units, yielding 
\beq
r_0^4 (\Delta_{55}^3-\Delta_{55}^\pm) = -0.10 \pm 0.06 \pm 0.03  \, ,
\label{D55SPLIT}
\eeq
where the first error is essentially statistical and stems from the evaluation of $\xi_\pi$ in 
eq.~(\ref{XIPI_DETER}) and the second is the systematic uncertainty reflecting the 
approximations involved in the way we parametrize (sect.~\ref{sec:C1}) and evaluate 
(sect.~\ref{sec:NUMERICO}) the quantity $\Delta_{55}^3-\Delta_{55}^\pm$. 

We see from Fig.~\ref{fig:figMPISPLIT} that within statistical errors (mainly coming from the 
noisy quark-disconnected diagrams entering the calculation of $m_{\pi^3}$) the available data 
for $\Delta m^2_{\pi}|_L$ show the expected scaling behaviour with the lattice spacing. 
The quark mass dependence of $(r_0/a)^2 r_0^2 \Delta m^2_{\pi}|_L$ turns out to be of 
reasonable magnitude and not inconsistent with our theoretical estimate~(\ref{zetapi_th}) within 
the large uncertainty affecting $(d \Delta m^2_{\pi} / d \hat\mu_q )|_L$ (from the plotted data 
one finds $(d \Delta m^2_{\pi} / d \hat\mu_q )|_L\sim 300 \div 900$~MeV at $\beta=3.9$). 

The estimated value of $r_0^4 (\Delta_{55}^3-\Delta_{55}^\pm)$ with its error (see 
eq.~(\ref{D55SPLIT})) is (in modulus) much smaller than the observed pion (squared) 
mass splitting, and also about twenty times smaller than $r_0^4 |\zeta_\pi|$. We thus 
conclude that the dominant contribution to $\Delta m^2_{\pi}|_L$ comes from the 
$a^2\zeta_\pi$ correction, as announced. 


\section{O($a^2$) isospin violating effects in Mtm-LQCD} 

In this section we want to provide numerical evidence of the fact that in Mtm-LQCD simulations
O($a^2$) isospin breaking effects are negligible in all the physical quantities measured up to
now by the ETM Collaboration, with the exception of the neutral pion mass.

In Table~\ref{tab:results} we report the relative difference of pseudoscalar meson squared masses
and decay constants between neutral and charged particles for pseudoscalar and
vector mesons, as well as the relative splitting between the masses of $\Delta^{+}$ and
$\Delta^{++}$ baryons. Most of these data already appeared in the conference
contributions of ref.~\cite{PROC} and in ref.~\cite{BARY}.

\label{sec:EXDA}

\begin{table}[htb]
  \centering
  \begin{tabular*}{1.\linewidth}{@{\extracolsep{\fill}}lccc}
    \hline\hline
    obs. ${\cal O}$ & $\beta$ & $a\mu_q$ & $R_s[{\cal O}]$ \\
    \hline\hline
    $m_\mathrm{PS}$ & $3.90$ & $0.0040$ & $0.185(44)$ \\
    $m_\mathrm{PS}$ & $3.90$ & $0.0085$ & $0.139(51)$ \\
    $m_\mathrm{PS}$ & $4.05$ & $0.0030$ & $0.120(110)$ \\
    $m_\mathrm{PS}$ & $4.05$ & $0.0060$ & $0.120(42)$ \\
    \hline\hline
    $f_\mathrm{PS}$ & $3.90$ & $0.0040$ & $0.04(6)$ \\
    $f_\mathrm{PS}$ & $3.90$ & $0.0085$ & $-0.09(8)$ \\
    $f_\mathrm{PS}$ & $4.05$ & $0.0030$ & $-0.03(6)$ \\
    $f_\mathrm{PS}$ & $4.05$ & $0.0060$ & $0.01(5)$ \\
    \hline
    $m_\mathrm{V}$  & $3.90$ & $0.0040$ & $0.022(68)$ \\
    $m_\mathrm{V}$  & $3.90$ & $0.0085$ & $0.021(44)$ \\
    $m_\mathrm{V}$  & $4.05$ & $0.0030$ & $-0.104(108)$ \\
    $m_\mathrm{V}$  & $4.05$ & $0.0060$ & $0.003(49)$ \\
    \hline
    $m_\mathrm{V} f_\mathrm{V}$  & $3.90$ & $0.0040$ & $-0.07(18)$ \\
    $m_\mathrm{V} f_\mathrm{V}$  & $3.90$ & $0.0085$ & $-0.01(11)$ \\
    $m_\mathrm{V} f_\mathrm{V}$  & $4.05$ & $0.0030$ & $-0.31(29)$ \\
    $m_\mathrm{V} f_\mathrm{V}$  & $4.05$ & $0.0060$ & $-0.12(13)$ \\
    \hline
    $m_\Delta$      & $3.90$ & $0.0040$ & $0.022(29)$  \\
    $m_\Delta$      & $3.90$ & $0.0060$ & $0.001(21)$ \\
    $m_\Delta$      & $3.90$ & $0.0085$ & $0.005(19)$  \\
    $m_\Delta$      & $3.90$ & $0.0100$ & $0.007(19)$ \\
    $m_\Delta$      & $4.05$ & $0.0030$ & $-0.004(45)$ \\
    $m_\Delta$      & $4.05$ & $0.0060$ & $0.004(17)$  \\
    $m_\Delta$      & $4.05$ & $0.0080$ & $0.000(18)$  \\
    $m_\Delta$      & $4.05$ & $0.0120$ & $0.007(16)$  \\
    \hline\hline
  \end{tabular*}
  \caption[tab:results]{
The ratio $R_s[{\cal O}]$ of eq.~(\ref{eq:Rofs}) for the observables ${\cal O}$ indicated in the 
first column at the simulation parameters specified in the second and third column (lattice size is 
about $L=2.1$~fm). The large statistical error on $R_s$ for mesonic observables at $\beta=4.05$, 
$a\mu_q=0.0030$ is due to limited statistics for the quark-disconnected contributions to the neutral 
meson correlators.} 
 \label{tab:results}
\end{table}

%

We refer to~\cite{ETMC_long} for a detailed description of the techniques used to compute
correlators, extract physical observables and perform the necessary error analysis. The results 
presented in this paper are based on the configurations generated by the ETM Collaboration using 
the formulation of lattice QCD at maximal twist (with $c_{\rm SW}=0$~\cite{SW}) and tree-level 
Symanzik improved gluon action. We show data coming from $\beta=3.9$~\cite{LET} and 
$\beta=4.05$~\cite{PROC,BARY} simulations at two, roughly matched (in physical units) 
values of the twisted mass parameter $\mu_q$ for each $\beta$. Again we stress that the 
computation of the mass and decay constant of neutral (pseudoscalar and vector) 
mesons involves the evaluation of quark-disconnected diagrams, which is the reason for the 
relatively larger errors affecting these quantities. In the case of $\Delta$-baryon 
masses (where no computation of quark-disconnected diagrams is required) we 
give results at four values of $\mu_q$ for $\beta=3.9$ and $\beta=4.05$.

We quote in Table~\ref{tab:results} the relative isospin splitting, $R_s$, measured 
between the observables $\mathcal{O}$ and $\mathcal{O'}$, namely 
\begin{equation} 
R_s[\mathcal{O}] = \frac{\mathcal{O}-\mathcal{O'}}{\mathcal{O}}\, ,
\label{eq:Rofs} 
\end{equation}
with $\mathcal{O}$ ($\mathcal{O'}$) denoting the quantity in the charged (neutral)
sector in the case of meson data and the $\Delta^{+,0}$-mass ($\Delta^{++,-}$-mass)
in the case of nucleons. We recall that, owing to the invariance of the Mtm-LQCD 
action under parity and $u \leftrightarrow d$ flavour exchange, $\Delta^{+}$ and 
$\Delta^{0}$ baryons, as well as $\Delta^{++}$ and the $\Delta^{-}$, 
are exactly mass-degenerate.

It should also be remarked that to calculate $R_s[\mathcal{O}]$ in the cases of the pseudoscalar and 
vector (isotriplet) meson decay constants use has been made of the appropriate renormalization 
constants ($Z_A$ and $Z_V$~\footnote{We recall that $Z_A$ and $Z_V$ are the renormalization 
constants of the $\chi$-basis bare operators $\bar\chi \gamma_\mu \gamma_5 \tau^b \chi$ and 
$\bar\chi \gamma_\mu \tau^b \chi$, respectively. For the reader convenience we also report  
the renormalization rules for the currents in Mtm-LQCD~\cite{FR1,Dim_etal_Z} in the physical basis. 
For the charged ($b=\pm$) axial and vector currents they read 
\beq
(\bar\psi \gamma_\mu \gamma_5 \tau^{\pm} \psi)_R = Z_V \bar\psi \gamma_\mu \gamma_5 \tau^{\pm} \psi \, ,
\, \quad (\bar\psi \gamma_\mu \tau^{\pm} \psi)_R = Z_A \bar\psi \gamma_\mu  \tau^{\pm} \psi \, ,
\eeq
while for their neutral ($b=3$) counterparts one has 
\beq
(\bar\psi \gamma_\mu \gamma_5 \tau^{3} \psi)_R = Z_A \bar\psi \gamma_\mu \gamma_5 \tau^{3} \psi  \, ,
\, \quad (\bar\psi \gamma_\mu \tau^{3} \psi)_R = Z_V \bar\psi \gamma_\mu  \tau^{3} \psi \, .
\eeq})  
which were taken from ref.~\cite{Dim_etal_Z}. The expressions of the pseudoscalar  
and vector meson decay constants in the physical quark basis read (no sum over $b$)
\begin{eqnarray}
& f_{{\rm PS}^b} = \frac{1}{m_{{\rm PS}^b}} \langle \Omega | 
(\bar\psi \gamma_0 \gamma_5 \frac{1}{2}\tau^b \psi)_R | {\rm PS}^b \rangle \, , \qquad b=\pm,3 \, ,
\label{FPS} \\
& f_{{\rm V}^b} = \frac{1}{m_{{\rm V}^b}} \langle \Omega | 
(\bar\psi \gamma_k \frac{1}{2}\tau^b \psi)_R | {\rm V}^b \rangle \, , \qquad b=\pm,3 \, ,
\label{FV} 
\end{eqnarray}
where the suffix $R$ denotes renormalized operators and $k$ is a spatial index~\footnote{
To be consistent with the H(4) invariance of the meson correlators the matrix elements involving charged 
($b=\pm$) or neutral ($b=3$) mesons are normalized in eqs.~(\ref{FPS}) and~(\ref{FV}) 
with the corresponding charged or neutral meson mass.}. 
Preliminary results on $f_{{\rm V}^b}$ have appeared in~\cite{BK}. 

The striking conclusion which emerges looking at Table~\ref{tab:results} is that 
among the many relative splittings reported there we observe a large and statistically 
significant value of $R_s$ {\em only} when pion masses are compared. For all other 
observables the splitting is consistent with zero within the quoted errors.

\section{Concluding remarks}   
\label{sec:CONC} 

In this paper we have proposed a theoretical explanation for the size of 
the O($a^2$) flavour violating cutoff effects responsible for the large 
splitting between neutral ($m_{\pi^3}$) and charged ($m_{\pi^\pm}$) pion 
mass seen in recent Mtm-LQCD dynamical simulations~\cite{LET}. 
We have identified the origin of the additive O($a^2$) cutoff effects appearing 
in $m_{\pi^3}^2$, which are absent in $m_{\pi^\pm}^2$ as the latter has discretization 
errors only of order $a^2\mu_q$, $a^4$ and higher. The magnitude of 
the O($a^2$) lattice artifacts in the neutral pion mass is controlled, in the language 
of the Symanzik expansion, by the continuum matrix element 
$ \langle\pi^3(\vec 0)| [ {\cal L}_6^{\rm Mtm}(0) - 
(1/2) \int d^4x {\cal L}_5^{\rm Mtm}(x) {\cal L}_5^{\rm Mtm}(0) ] |\pi^3(\vec 0)\rangle|_{\rm cont} $,
which, if Mtm-LQCD is defined via the optimal critical mass estimate, is numerically dominated by
$\zeta_\pi=\langle\pi^3(\vec 0)|{\cal L}_6^{\rm Mtm}|\pi^3(\vec 0)\rangle|_{\rm cont}$. 
In the vacuum saturation approximation one gets the theoretical
estimate $|\zeta_\pi| \sim 2\hat{G}_\pi^2\sim (560 \div 580~{\rm MeV})^4$, 
up to an unknown multiplicative factor depending on the details of the lattice action 
and the number of dynamical quark flavours (actually one can argue 
that this coefficient grows linearly with $N_f$). 

We have shown that numerical simulations of Mtm-LQCD with $N_f=2$ light
dynamical quarks yield values of $(r_0/a^2)^2 r_0^2 \Delta m^2_{\pi}|_L$ with
negative sign and a magnitude that is in line with the above theoretical estimate.
We have also provided evidence that among the many physical quantities recently 
measured in numerical simulations only the neutral pion mass appears to be 
affected by large flavour-breaking discretization errors.  

Given the independence on the twist angle of the Symanzik effective action in the limit of 
vanishing quark mass, these results may have a bearing also on LQCD simulations  
employing standard (i.e.\ untwisted) Wilson fermions. Indeed also in this formulation O$(a^2$) 
cutoff effects of the same nature as those observed in Mtm-LQCD are expected to be present. In 
particular a formula completely analogous to eq.~(\ref{MP3}) holds true for the squared pion mass.  
Typically these O$(a^2$) cutoff effects are reabsorbed in the definition of the current quark mass 
(which by construction vanishes simultaneously with the pion mass) and thus in the corresponding 
definition of the critical mass. The consequences of this way of proceeding are in principle detectable 
in the form of O($a^2$) cutoff effects in the masses of all the other hadrons and, more generally, 
in all the quantities that have a non-negligibly small dependence on the quark mass. Actually, in what 
observables exactly they will appear depends on the precise way the critical mass happens to have 
been determined (see the analysis in Appendix~A).  

\appendix 
\renewcommand{\thesection}{A} 
\section*{Appendix A - Symanzik expansion and critical mass in LQCD with Wilson fermions} 
\label{sec:APPA} 

We consider $N_f=2$ LQCD with quarks regularized as Wilson fermions. 
For generic values of the bare (twisted, $\mu_q$, and untwisted, $m_0$) 
mass parameters the lattice action reads~\footnote{We adopt the 
conventions and notations of refs.~\cite{FR1,FR2,FR3}. In particular the quark 
fields in the ``physical'' and ``twisted'' basis will be denoted by $\psi,\bar\psi$ 
and $\chi,\bar\chi$, respectively.}  
\beqn
S_L=S^{\rm YM}_L+\bar\chi\left[\gamma\cdot\widetilde{\nabla}-\frac{a}{2}\nabla^*\nabla
+c_{\rm SW}\frac{ia}{4}\sigma\cdot F+m_0+i\mu_q\gamma_5\tau^3\right]\chi\, ,
\label{LATACT}
\eeqn
where $\widetilde{\nabla}_\mu = \frac{1}{2} [ \nabla_\mu + \nabla^*_\mu]$, with
$\nabla_\mu$ ($\nabla^*_\mu$) the forward (backward) gauge covariant lattice derivative. 
For the sake of generality we have also allowed for the presence of the clover term. We are especially 
interested in two specific regularizations of the Dirac-Wilson action comprised in~(\ref{LATACT}).

The first is the Mtm-LQCD regularization which is obtained from~(\ref{LATACT}) by setting 
$\mu_q= {\rm O}(a^0)$ and $m_0 = M_{\rm cr}^{e}$, where $M_{\rm cr}^e$ is some 
estimate of the critical mass. The physical interpretation of this scheme is most transparent in 
the quark basis resulting from the field transformation 
\beqn
&&\hspace{-.2cm}\psi=\exp(i\pi\gamma_5\tau^3/4)\chi\, ,\quad 
\bar\psi=\bar\chi\exp(i\pi\gamma_5\tau^3/4)\, ,
\label{CTPB}\eeqn
where the lattice action takes the form
\beqn
\hspace{-.2cm}S_L^{\rm Mtm} = S^{\rm YM}_L + 
\bar\psi \left[\gamma \!\cdot\! \widetilde{\nabla} -i\gamma_5\tau^3 
\left(- \frac{a}{2} \nabla^*\nabla 
+c_{\rm SW}\frac{ia}{4}\sigma\!\cdot\!F+M_{\rm cr}^{e}\right)+\mu_q\right]\psi\, .
\label{LATACTMTM}
\eeqn 
This quark basis is referred to as the ``physical'' basis, because the quark mass term takes 
the form $\mu_q\bar\psi\psi$ with $\mu_q$ real~\cite{FR1}.

The second is the (clover) standard Wilson fermion action, $S_L^{\rm cl}$, which is 
obtained by setting $\mu_q=0$ and $m_0 = m+M_{\rm cr}^{e}$ with $m$ an 
O$(a^0)$ quantity. With this choice the most appropriate basis for discussing physics is the 
$\chi$-basis itself in which eq.~(\ref{LATACT}) was written in the first place.

\subsection{Symanzik expansion of Wilson fermion LQCD}
\label{sec:SYMW}

The Symanzik effective Lagrangian associated to the Wilson action~(\ref{LATACT}) reads 
\beqn 
&&{\cal L}_{\rm Sym} = {\cal L}_4 + \delta {\cal L}_{\rm Sym} \, ,\label{LELgen}\\
&&{\cal L}_4 = {\cal L}^{\rm YM} +\bar\chi[D\hspace{-0.3cm}/ +m+i\gamma_5\tau^3\mu_q]\chi \, ,\label{L4gen}\\
&&\delta {\cal L}_{\rm Sym} = a {\cal L}_5 + a^2 {\cal L}_6 + {\rm O}(a^3) \, ,\label{L56} 
\eeqn 
where the four-dimensional operator, ${\cal L}_4$, specifies the formal target  
theory in which continuum correlators are evaluated. The very definition of effective action
(all the necessary logarithmic factors are understood) as a tool to describe the $a$ 
dependence of lattice correlators implies that the mass parameters in ${\cal L}_4$, 
if not exactly vanishing, must be O($a^0$) quantities. Thus all the lattice artifacts 
affecting $M_{\rm cr}^e$ will be described by higher dimensional operators 
in $\delta {\cal L}_{\rm Sym}$ of the form 
$a^{k}\delta_k \Lambda_{\rm QCD}^{k+1}\bar\chi\chi$, $k\geq 1$, with 
$\delta_k$ dimensionless coefficients, which are functions of the gauge coupling. 

After using the equations of motion entailed by ${\cal L}_4$, the O($a$) piece of 
$\delta {\cal L}_{\rm Sym}$ reads~\cite{LSSW,FMPR} 
\beq 
{\cal L}_5 = b_{5;{\rm SW}} \bar\chi i \sigma \cdot F \chi + 
\delta_1 \Lambda_{\rm QCD}^2\bar\chi \chi + {\rm O}(m,\mu_q) \, .
\label{L5gen} 
\eeq  
The terms multiplied by powers of $m$ and/or $\mu_q$ are not specified in eq.~(\ref{L5gen}) 
as they are not of relevance for the topic discussed in this paper. We recall that the 
coefficients $b_{5;{\rm SW}}$ and $\delta_1$ vanish if $c_{\rm SW}$ in eq.~(\ref{LATACT}) 
is set to the value appropriate for Symanzik O($a$) improvement. 

The O($a^2$) part of $\delta {\cal L}_{\rm Sym}$ has a more complicated expression, of the type 
\beq 
{\cal L}_6=\sum_{i=1}^3 b_{6;i}\Phi_{6;i}^{\rm glue}+b_{6;4}\bar\chi\gamma_\mu (D_\mu)^3\chi+ 
\sum_{i=5}^{14} b_{6;i} \Phi_{6;i}+\delta_2 \Lambda_{\rm QCD}^3 \bar\chi\chi +{\rm O}(m,\mu_q)\, ,
\label{L6gen} 
\eeq 
where the first three operators are purely gluonic, the fourth is a non-Lorentz  
invariant fermionic bilinear and the remaining ones are four-fermion operators, 
which we choose to write in the form (equivalence with the list in~\cite{SW} 
can be proved by using Fierz rearrangement)
\beq\begin{array}{lll}
& \hspace{-.8cm}\Phi_{6;5}  =\frac{1}{4} (\bar\chi \chi) (\bar\chi \chi) \, , \quad  
& \Phi_{6;6}  = \frac{1}{4}\sum_b(\bar\chi \tau^b \chi) (\bar\chi \tau^b \chi) \, ,  
\nonumber \\ 
& \hspace{-.8cm}\Phi_{6;7}  = \frac{1}{4}(\bar\chi i\gamma_5 \chi) (\bar\chi \gamma_5 \chi) \, , \quad  
& \Phi_{6;8}  = \frac{1}{4}\sum_b(\bar\chi i \gamma_5 \tau^b \chi) (\bar\chi \gamma_5 \tau^b \chi) \, ,  
\nonumber \\ 
&\hspace{-.8cm}\Phi_{6;9}  =\frac{1}{4} (\bar\chi \gamma_\lambda \chi) (\bar\chi \gamma_\lambda \chi) \, , \quad  
& \Phi_{6;10} = \frac{1}{4}\sum_b(\bar\chi \gamma_\lambda \tau^b \chi) (\bar\chi \gamma_\lambda \tau^b \chi) \, ,  
\nonumber \\ 
&\hspace{-.8cm} \Phi_{6;11} = \frac{1}{4}(\bar\chi \gamma_\lambda \gamma_5 \chi)  
               (\bar\chi \gamma_\lambda \gamma_5 \chi) \, , \quad  
& \Phi_{6;12} = \frac{1}{4}\sum_b(\bar\chi \gamma_\lambda \gamma_5 \tau^b \chi)  
               (\bar\chi \gamma_\lambda \gamma_5 \tau^b \chi) \, ,  
\nonumber \\ 
& \hspace{-.8cm}\Phi_{6;13} = \frac{1}{4}(\bar\chi \sigma_{\lambda\nu} \chi)  
               (\bar\chi \sigma_{\lambda\nu} \chi) \, , \quad  
& \Phi_{6;14} = \frac{1}{4}\sum_b(\bar\chi \sigma_{\lambda\nu} \tau^b \chi)  
               (\bar\chi \sigma_{\lambda\nu} \tau^b \chi) \, . \label{4FinL6} 
\end{array} 
\eeq
In closing this section it is important to note that the form of the Symanzik effective 
Lagrangian enjoys some interesting degree of universality in the sense that formally 
its zero mass limit ($m_0-M_{\rm cr}^e=\mu_q=0$) only depends on discretization 
details, like the form of the gauge action, the specific expression of the lattice 
derivatives, or the value of $c_{\rm SW}$, but not on whether one shall be 
eventually dealing with standard or twisted Wilson fermions. 

When quark mass terms are switched on, the QCD vacuum gets polarized driving 
spontaneous chiral symmetry breaking. The information about spontaneous chiral 
symmetry breaking effects is embodied in the Symanzik analysis by assigning to 
the continuum correlators in the expansion appropriate values consistent with 
the residual exact symmetries of the target QCD theory. In this way the relative 
``orientation'' of the quark mass term with respect to the explicitly chirally breaking terms 
of the Symanzik low energy Lagrangian (describing the effects of the Wilson term present 
in the lattice action) becomes crucial for determining the size and the order in $a$ of cutoff
artifacts in correlation functions~\cite{FR1,FMPR}. 

\subsection{Critical mass in Wilson fermion LQCD}
\label{sec:CMW}

Both in the case of standard ($\mu_q=0$) and twisted mass ($\mu_q={\rm O}(a^0)$) 
Wilson fermions the critical mass is taken as the value of $m_0$ at which the PCAC mass 
vanishes. It is, however, useful to examine separately the two cases, 
since the resulting structure of lattice artifacts affecting these two determinations of the 
critical mass will be significantly different, as alluded to at the end of the previous section. 

\subsubsection{Critical mass in twisted-mass LQCD}
\label{sec:CMTM}

In twisted-mass LQCD the condition (which we write in the physical $\psi$-basis~(\ref{CTPB}))
\beq
a^3 \sum_{\vec x} \langle (\bar\psi \gamma_0 \tau^2 \psi)(\vec x,t)
(\bar\psi \gamma_5 \tau^1 \psi)(0) \rangle\Big{|}_L =0 
\label{LATCOR}
\eeq
leads to a determination of the critical mass which is ``optimal'' 
($M_{\rm cr}^{\rm opt}$) in the sense that with this choice 
all the leading chirally enhanced cutoff effects (i.e.\ those of relative 
order $(a/\mu_q)^{2k}$, $k=1,2,\ldots$ with respect to the dominant  
term in the $a\to 0$ limit) are eliminated from lattice correlators~\cite{FMPR}. 

In the spirit of the Symanzik expansion the condition~(\ref{LATCOR}) must be viewed as 
a relation holding true parametrically for generic values of $a$ (and $\mu_q$). As a 
consequence it is equivalent to an infinite set of equations, where each equation results 
from the vanishing of the coefficient of the term proportional to $a^k$, $k=0,1,2,\dots$, 
in the expansion of the l.h.s.\ of~(\ref{LATCOR})~\footnote{The practical problems 
in determining $m_0$ from~(\ref{LATCOR}), which are related to subtleties associated 
to the exchange of continuum ($a\to 0$) and chiral ($\mu_q\to 0$) limit and to the
possible limitations of the validity of the Symanzik expansion of the 
correlator~(\ref{LATCOR}), have been discussed at length in refs.~\cite{FMPR,SSS} 
and will not be repeated here.}.

That the condition~(\ref{LATCOR}) yields an acceptable estimate of the critical mass 
can be inferred from the observation that the first of these equations (the one which 
corresponds to the vanishing of the $a^0$ term) is nothing but the continuum relation 
\beq
\int d^3x\, \langle 
(\bar\psi \gamma_0 \tau^2 \psi)(\vec x,t)
(\bar\psi \gamma_5 \tau^1 \psi)(0) \rangle\Big{|}_{\rm cont} =0\, , 
\label{LATCORC}
\eeq
by which restoration of parity and isospin is enforced in the lattice theory. This means that, if (the O($a^0$) term in) 
$m_0$ is chosen so as to fulfill eq.~(\ref{LATCOR}), then we will simultaneously have $m=0$ in 
eq.~(\ref{L4gen}) and the identification on the lattice of the operator $\bar\psi\gamma_0\tau^2\psi$
with the time component of the vector current, $V_0^2$. At this point to more clearly understand the 
further implications of eq.~(\ref{LATCOR}), it is convenient to explicitly write down the first few 
terms of its Symanzik expansion for which one gets
\beqn
\hspace{-.5cm}0&=&-a\int d^3x\int d^4y\langle{\cal L}_5^{\rm Mtm}(y)V_0^2(x) P^1(0)\rangle\Big{|}_{\rm cont}
  \nonumber \\
\hspace{-.5cm}&+&a\int d^3x\langle [\Delta_1 V_0^2(x) P^1(0)+V_0^2(x)\Delta_1 P^1(0)]\rangle\Big{|}_{\rm cont}
  \nonumber \\
\hspace{-.5cm}&-&a^2\int d^3x\int d^4y\langle{\cal L}_6^{\rm Mtm}(y)V_0^2(x) P^1(0)\rangle\Big{|}_{\rm cont} 
  \nonumber \\
\hspace{-.5cm}&+&a^2\int d^3x\langle [\Delta_2 V^2_0(x) P^1(0)+V_0^2(x)\Delta_2 P^1(0)]\rangle\Big{|}_{\rm cont}
+ {\rm O}(a^3) \, .
\label{SYMCO}
\eeqn 
Eq.~(\ref{SYMCO}) has been written under the assumption that the O($a^0$) piece of 
$m_0$ has been chosen so that condition~(\ref{LATCORC}) is fulfilled. As a consequence 
(see eq.~(\ref{L5gen})) ${\cal L}_5^{\rm Mtm}$ is a parity-odd and flavour breaking 
operator of the form
\beqn
{\cal L}_5^{\rm Mtm} = b_{5;SW} \bar\psi \gamma_5\tau^3 \sigma \cdot F \psi 
- \delta_1 \Lambda_{\rm QCD}^2\bar\psi i \gamma_5 \tau^3 \psi + {\rm O}(\mu_q) \, ,
\label{L5Mtm}
\eeqn
while ${\cal L}_6^{\rm Mtm}$ (see eq.~(\ref{L6gen})) can be split into the sum of two contributions 
\beq
{\cal L}_6^{\rm Mtm}={\cal L}_6^{\rm P-even} 
- \delta_2\Lambda_{\rm QCD}^3\bar\psi i\gamma_5\tau^3\psi+{\rm O}(\mu_q) \, ,\label{L6}
\eeq
with ${\cal L}_6^{\rm P-even}$ a parity-even operator. With the notation 
$\Delta_j O$ in eq.~(\ref{SYMCO}) we indicate the operators of dimension $d_O+j$ that 
correct to O($a^j$) the (local) operator $O$. Their parity is equal to $(-1)^j$ times the 
parity of $O$ (see Appendix in~\cite{FMPR}). We explicitly remark that in 
eq.~(\ref{SYMCO}) we have omitted the O($a^2$) terms with two integrated insertions 
of ${\cal L}_5^{\rm Mtm}$ and the associated contact terms, because they all vanish 
by continuum parity. Finally for the purpose of the present argument we have 
ignored operators having coefficients proportional to $\mu_q$. 

At O($a$) the condition implied by eq.~(\ref{SYMCO}) reduces to~\cite{FMPR} 
\beqn
\hspace{-1.cm}&0& = \int d^3x \int d^4y \langle
[b_{5;SW} \bar\psi \gamma_5\tau^3 \sigma \cdot F \psi   \nn\\
&-& \delta_1  \Lambda_{\rm QCD}^2\bar\psi i \gamma_5 \tau^3 \psi ](y)
V_0^2(x) P^1(0) \rangle\Big{|}_{\rm cont}+{\rm O}(\mu_q) \, ,
\label{OACOND5}
\eeqn
since the first term in the second line of eq.~(\ref{SYMCO}) gives a vanishing 
contribution to eq.~(\ref{OACOND5}) in the chiral limit and there are no O($a$) 
operator corrections to $P^1$ (i.e.\ $\Delta_1 P^1=0$).

In the chiral regime, where intermediate pion states dominate, one recognizes that 
eq.~(\ref{OACOND5}) leads to the condition which determines the O($a$) piece of 
the optimal critical mass. We recall that with the definition
\beq
\xi_\pi=\xi_\pi(\mu_q)=\langle\Omega|\sum_{\ell=0}^{\infty}a^{2\ell} 
{\cal L}_{2\ell +5}^{\rm Mtm}|\pi^3(\vec 0)\rangle\Big{|}_{\rm cont}\, ,
\label{DEFXI}\eeq 
which to leading order in $a$ reads $\xi_\pi \simeq \langle\Omega|{\cal L}_{5}^{\rm Mtm}|\pi^3(\vec 0)\rangle |_{\rm cont}$, 
eq.~(\ref{OACOND5}) becomes, in the chiral regime $\mu_q \ll \Lambda_{\rm QCD}$, the constraint~\cite{FMPR} 
\beq
\xi_\pi = \langle\Omega|{\cal L}_5^{\rm Mtm}|\pi^3(\vec 0)\rangle\Big{|}_{\rm cont} 
\, + \, {\rm O}(a^2) = {\rm O}(\mu_q) \, + \, {\rm O}(a^2) \, ,
\label{OCM}
\eeq
which again should be looked at as an infinite set of equations, one for each power of $a$,
with the leading one fixing $\delta_1$ in terms of $b_{5;SW}$.

At O($a^2$) since by continuum parity invariance one can derive the equation 
\beqn
&&\int d^3x \int d^4y \langle{\cal L}_6^{\rm P-even}(y)
V_0^2(x) P^1(0) \rangle\Big{|}_{\rm cont}=0 \, ,\label{OACOND6}\\
&&\int d^3x\langle [\Delta_2 V^2_0(x) P^1(0)+V_0^2(x)\Delta_2 P^1(0)]\rangle\Big{|}_{\rm cont}=0\, ,
\label{OACOND7}
\eeqn
while the terms with the insertion of $\bar\psi i\gamma_5\tau^3\psi$ do not vanish 
\beq
\int d^3x\int d^4y\langle \bar\psi i\gamma_5\tau^3\psi(y)
V_0^2(x)P^1(0)\rangle\Big{|}_{\rm cont}\neq 0\, ,
\label{NEZ}
\eeq
the condition implied by eq.~(\ref{LATCOR}) (or eq.~(\ref{SYMCO})) 
yields $\delta_2=0$ in~(\ref{L6}). The important consequence of this argument is that the estimate of the 
critical mass provided by~(\ref{LATCOR}) is not affected by O($a^2$) corrections. 

The above line of reasoning can be generalized to all orders in $a$ leading to the conclusion that, 
if the critical mass is determined in twisted-mass LQCD by means of the condition~(\ref{LATCOR}) 
(whose Symanzik expansion takes the form~(\ref{SYMCO})),  
the expansion of the critical mass will only display O($a^{2p+1}$), $p=0,1,\ldots$, lattice 
corrections, with coefficients $\delta_{2p+1}$ determined by constraints analogous to~(\ref{OACOND5}).
 
\subsubsection{Critical mass in standard Wilson fermion LQCD}
\label{sec:CMSCW}

In the case of standard (clover) Wilson fermions the condition for the vanishing 
of the PCAC mass is ($b=1,2,3$, no sum over $b$) 
\beq
\frac{\tilde\partial_0\sum_{\vec x}\langle A_0^b({\vec x,t})P^b(0)\rangle}
{2\sum_{\vec x}\langle P^b({\vec x,t})P^b(0)\rangle}
\Big{|}_L \equiv m_{\rm PCAC}\Big{|}_L=0\quad \mathrm{for}\;\; \,\mu_q=0\, ,
\label{MPCAC}\eeq
which is analogous to eq.~(\ref{LATCOR}) with the crucial difference that 
now $\mu_q = 0$. The condition~(\ref{MPCAC}) is in practice implemented by 
looking for the limiting value of $m_0$ at which $m_{\rm PCAC}\to 0^+$. We will 
call $M_{\rm cr}^{e_W}$ the estimate of the critical mass obtained in this 
way~\footnote{This non-perturbative estimate of the critical mass is 
the one that is implicitly adopted in the studies of standard (clover) Wilson LQCD where  
the renormalized quark mass is defined as $\hat{m} = Z_A Z_P^{-1} m_{\rm PCAC}$.}.
 
Setting $m_0=M_{\rm cr}^{e_W}$ in the standard Wilson action corresponds to have 
the continuum mass in ${\cal L}_4$ equal to zero. At the same time lattice artifacts 
of order $a^k$, $k=2,3,...$ in $M_{\rm cr}^{e_W}$ will be correspondingly described 
by terms of the kind $a^k\delta_k^{e_W}\Lambda_{\rm QCD}^{k+1}\bar\chi\chi$ in ${\cal L}_k$. 
The O($a$) term is absent, if the lattice theory is clover improved. For the rest, unlike what 
we have shown to happen in twisted-mass LQCD (see sect.~\ref{sec:CMTM}), discretization 
errors of any order in $a$ will affect the critical mass determination provided by eq.~(\ref{MPCAC}).

As in the case of twisted-mass LQCD, also here one must look at the conditions coming from the 
vanishing of the PCAC mass as equations that fix the values of the coefficients $\delta_k^{e_W}$ 
in terms of the matrix elements of ${\cal L}_k$ between one-pion states. For instance, at 
O($a^2$) from the symmetries of the standard Wilson action (and the form of the 
associated Symanzik expansion) one cannot conclude anymore that $\delta_2^{e_W}$ vanishes. 
Rather the value of this parameter is fixed through eq.~(\ref{MPCAC}) by the condition 
\beq
\langle\pi(\vec 0)|{\cal L}_6|\pi(\vec 0)\rangle\Big{|}_{\rm cont}=0\, ,
\label{SWC}\eeq
where ${\cal L}_6$ is the full six-dimensional operator of the 
Symanzik effective Lagrangian (given in eq.~(\ref{L6gen})) which, we recall, also 
includes the $\delta_2^{e_W}\Lambda_{\rm QCD}^3 \bar\chi\chi$ term. Eq.~(\ref{SWC}) 
follows from the fact that the lattice pion squared mass and $m_{\rm PCAC}|_L$
by construction vanish at the same value of $m_0$. Furthermore these two quantities are found 
to be linearly proportional to a very good approximation in the region where they are both small, 
implying that possible additive cutoff artifacts affecting  
$m_{\rm PCAC}|_L$ and the squared pion mass are also proportional to each other.  

\appendix 
\renewcommand{\thesection}{B} 
\section*{Appendix B - O($a^2$) corrections to the squared pion mass}
\label{sec:APPB} 

The correlator of interest for extracting the squared pion mass is the 
four-dimensional Fourier transform of the two-point subtracted~\footnote{Since we are
interested in determining the structure of the cutoff corrections affecting the lattice pion 
mass, we can always imagine that contact terms, which do not display the pion pole, have 
been subtracted out. In the formulae that follow the superscript ``$^{subtracted}$'' is thus 
always understood.} pseudoscalar correlator (no sum over $b=\pm,3$) which reads 
\beq
\Gamma^b_L(p)=a^4\sum_x e^{ipx}\langle P^b(x)P^b(0)\rangle\Big{|}_L\, .
\label{P3P3}
\eeq
$\Gamma^b_L(p)$ has a pole at $p^2=-m^2_{\pi^b}|_L$ with a residue given by 
$|G_{\pi^b}|_L^2$, where (see eq.~(\ref{G-SIGMA-DEF}))
\beq
G_{\pi^b}\Big{|}_L=\langle\Omega|P^b|\pi^b(\vec 0)\rangle\Big{|}_L\, .\label{GP3A}
\eeq
To proceed further we need to write down the Symanzik expansion of $\Gamma_L(p)$ up to 
order $a^2$ included. This gives 
\beqn
\hspace{-1.cm} &&a^4\sum_{x}  e^{ipx} \langle P^b({x})P^b(0)\rangle\Big{|}_L = 
\int d^4{x}\,  e^{ipx} \langle P^b({x})P^b(0)\rangle\Big{|}_{\rm cont} \nn\\
\hspace{-1.cm}&+&a^2\int d^4{x}\,  e^{ipx} \langle \Delta_1 P^b({x})
\Delta_1 P^b(0)\rangle\Big{|}_{\rm cont} 
\nn\\\hspace{-1.cm}&+&
a^2\int d^4{x}\,  e^{ipx} \langle \Delta_1 P^b({x})P^b(0)
\int d^4y\,{\cal L}_5^{\rm Mtm}(y)\rangle\Big{|}_{\rm cont} \nn\\
\hspace{-1.cm}&+&a^2\int d^4{x}\,  e^{ipx} \langle P^b({x})\Delta_1 P^b(0)
\int d^4y\,{\cal L}_5^{\rm Mtm}(y)\rangle\Big{|}_{\rm cont}+\nn\\
\hspace{-1.cm}&+&\frac{a^2}{2}\int d^4{x}\,  e^{ipx} \langle P^b({x})P^b(0)
\int d^4y\,{\cal L}_5^{\rm Mtm}(y)\int d^4y'\,
{\cal L}_5^{\rm Mtm}(y')\rangle\Big{|}_{\rm cont} \nn\\
\hspace{-1.cm}&-&a^2\int d^4{x}\,  e^{ipx} \langle \Delta_2 P^b({x})P^b(0)\rangle\Big{|}_{\rm cont}-
a^2\int d^4{x}\,  e^{ipx} \langle P^b({x})\Delta_2 P^b(0)\rangle\Big{|}_{\rm cont} \nn\\
\hspace{-1.cm}&-&a^2\int d^4{x}\,  e^{ipx} \langle P^b({x})P^b(0)
\int d^4y\,{\cal L}_6^{\rm Mtm}(y)\rangle\Big{|}_{\rm cont}+{\rm O}(a^4)\, .
\label{WTINP}\eeqn
The absence of odd powers of the lattice spacing in this formula is guaranteed 
by the results of ref.~\cite{FR1}, as we are working at maximal twist. We will 
also assume that the critical mass has been determined in the optimal way described 
in sect.~\ref{sec:CMTM}, so that by setting $m_0=M_{\rm cr}^{\rm opt}$ the 
condition~(\ref{OCM}) holds true. We also recall that the operators $\Delta_1 P^b$ 
and $\Delta_2 P^b$ are the dimension four and five terms that are needed to 
compensate for the O($a$) and O($a^2$) operator corrections arising from the 
contact of ${\cal L}_5^{\rm Mtm}(y)$ and ${\cal L}_6^{\rm Mtm}(y)$, respectively, 
with the operators localized at the points $x$ and 0.

Since we are interested in the lattice corrections to the pion mass, we must 
look in the r.h.s.\ of eq.~(\ref{WTINP}) for the terms which may display 
the double pion pole factor~\footnote{For short from now on we will simply write $m_\pi^2$ 
(with no isospin index as in the continuum target theory all three pions are mass-degenerate) 
for $m_\pi^2|_{\rm cont}$.} $(p^2 + m_\pi^2)^{-2}$.

Among the many terms in the r.h.s.\ of eq.~(\ref{WTINP}) only the fifth and the 
last display the double pole we are after. Putting together eqs.~(\ref{G-SIGMA-DEF}), 
(\ref{GP3A}) and~(\ref{WTINP}) and taking the limit $p \to 0$ at small $m_\pi^2$ (but 
always parametrically larger than $a^2$~\cite{FMPR}), from the ensuing pion pole 
dominance we arrive at the formula 
\beqn
\frac{|G_{\pi^3}|^2}{m^2_{\pi^3}}\Big{|}_L=\frac{|G_\pi|^2}{m^2_{\pi}}\Big{|}_{\rm cont}\Big{[}1 -
a^2\frac{\langle\pi^3(\vec 0)|{\cal L}_{6\, \& 55}^{\rm Mtm}(0)|\pi^3(\vec 0)\rangle}
{m_{\pi}^2}\Big{|}_{\rm cont}\Big{]}+{\rm O}(\frac{a^2}{m_{\pi}^2})\, ,
\label{NPM}\eeqn
with 
\beq
{\cal L}_{6\, \& 55}^{\rm Mtm}(0)={\cal L}_6^{\rm Mtm}(0) - \half \int d^4 x {\cal L}_5^{\rm Mtm}(x)
{\cal L}_5^{\rm Mtm}(0)\, .
\label{D56}
\eeq
A simple Taylor series resummation leads precisely to eq.~(\ref{MP3}) of the text, which we stress 
also agrees with the results derived in $\chi$PT~\cite{SCOR,MUSC,SHWU,SSS}. 

We close the Appendix by noting that the absence of a contribution from 
$\langle\pi^\pm(\vec 0)|{\cal L}_6^{\rm Mtm}(0)|\pi^\pm(\vec 0)\rangle|_{\rm cont}$ 
in eq.~(\ref{DMPPM}) is a consequence of the commutation relation~\footnote{In Mtm-LQCD the 
flavour symmetry group is the ``oblique'' SU(2)$_{\rm ob}$ group generated by the charges 
$Q_A^+$, $Q_A^-$ and $Q_V^3$. These conserved charged are associated with the (one-point 
split) currents reported in sect.~4.1 of ref.~\cite{FR1}. In the notation of that paper we are dealing 
here with the case $r=1$, $\omega_r=\pi/2$.} 
\beq
[Q_A^\pm,{\cal L}^{\rm Mtm}_{\rm Sym}]={\rm O}(\mu_q)={\rm O}({m^2_{\pi}})\, ,
\label{CR}
\eeq
from which one gets the SPT 
\beqn
\hspace{-1.2cm}&&\langle\pi^\pm(\vec 0)|{\cal L}^{\rm Mtm}_{6}(0)
|\pi^\pm(\vec 0)\rangle\Big{|}_{\rm cont}=\nn\\
\hspace{-1.2cm}&&=\frac{-i}{f_\pi}\langle\Omega|[Q_A^\pm,{\cal L}^{\rm Mtm}_{6}(0)]
|\pi^\pm(\vec 0)\rangle\Big{|}_{\rm cont}+{\rm O}({m^2_\pi})={\rm O}({m^2_\pi})\, .
\label{OSPTC}\eeqn


\appendix 
\renewcommand{\thesection}{C} 
\section*{Appendix C - Expression and estimates of $\Delta^{\pm,3}_{55}$}
\label{sec:APPC} 

In this (long) Appendix we present arguments showing that $\Delta^b_{55}$, $b=3,\pm$ (eq.~(\ref{D55})) 
yield contributions to $m^2_{\pi^b}|_L$ (eqs.~(\ref{DMPPM}) and~(\ref{DMP3})) which 
are parametrically (in the limit $m_\pi^2\to 0$) and/or numerically negligible, provided Mtm-LQCD is 
defined by setting $m_0$ (see eq.~(\ref{LATACT})) at its optimal critical value determined by the 
condition~(\ref{LATCOR}). 

In sect.~\ref{sec:C1} we provide the expression of $\Delta^{3}_{55}$ and $\Delta^{\pm}_{55}$ (eq.~(\ref{D55}))
based on standard field theoretical manipulations, i.e.\ insertion of intermediate states, or alternatively LSZ representation 
and reduction formulae, and, wherever applicable, SPT's~\cite{CCW,WEIN}. In sect.~\ref{sec:NUMERICO} we give 
a numerical estimate of the magnitude of $\Delta^{3}_{55}$ and $\Delta^{\pm}_{55}$ for the typical 
lattice volume and pion mass values relevant for the recent ETMC simulations. The details of our estimate 
of the key matrix element $\xi_\pi$ (eqs.~(\ref{DEFXI}) and~(\ref{OCM})), which controls the numerically dominant 
(in the small $m_\pi^2$ region) contributions to $\Delta^{3,\pm}_{55}$, are deferred to sect.~\ref{sec:PROP}.  

\subsection{The theoretical analysis of $\Delta^{3}_{55}$  and $\Delta^{\pm}_{55}$}
\label{sec:C1} 

For the purposes of our study we restrict attention to $N_f=2$ QCD with light $u$ and $d$ quarks 
yielding light, weakly interacting pions. In this regime, on general grounds, one can write 
in the infinite volume limit the spectral decomposition of the continuum quantity $\Delta^3_{55}$ 
(eq.~(\ref{D55})) in the convenient form~\footnote{For short in this Appendix we will drop the 
subscript $|_{\rm cont}$ from continuum QCD quantities. Lattice quantities will still be labeled by 
the subscript $|_{L}$.} 
\beqn
&&\hspace{-0.9cm} \Delta^3_{55}= - \frac{\xi_\pi}{m^2_\pi}
\langle\pi^3(\vec 0)|{\cal L}_5^{\rm Mtm}|\pi^3(\vec 0)\pi^3(\vec 0)\rangle 
+ \label{D55neu}  \\
&&- \frac{4\pi}{16\pi^3} \int_{2 m_\pi}^{4m_\pi} dE \frac{k(E) }{E^2 - m^2_\pi}
  | \langle\pi^3(\vec 0)|{\cal L}_5^{\rm Mtm} | {\cal P}_{\pi\pi}^{(I_3=0)} 
    \pi \pi, E \rangle |^2 + R_{55}^{3}  \, , \nonumber \\
&&\hspace{-0.9cm} R_{55}^{3}=- \sum_{n} \frac{ |\langle\pi^3(\vec 0)|{\cal L}_5^{\rm Mtm}|
        \sigma^0_n(\vec 0)\rangle|^2 }{m_{\sigma^0_n}^2 - m^2_\pi} +\ldots \, , \label{REMNEU}
\eeqn
where we have explicitly separated out from the remainder, $R_{55}^3$, the contribution of
the semi-disconnected one-pion pole (sometimes referred to as ``tadpole'' in the following) 
and the contribution of the cut over two-pion states below the inelastic threshold. In order 
to have more manageable formulae we have decided to reduce the initial integrations over the 
momenta of the two pions to the integration to the single center of mass energy variable, and we have set 
\beq 
k(E) = \sqrt{(E/2)^2-m_\pi^2}\, .
\label{KAP}\eeq 
The remainder in eq.~(\ref{REMNEU}) comprises a sum over one-particle scalar states with zero 
isospin~\footnote{Phenomenologically the lightest of these states should be identified with the 
$a_0(980)$ meson. It should be kept in mind that, since in the actual simulation data we will 
be considering the lightest ``pion'' mass is about 300~MeV, the lattice state corresponding to the 
``$a_0(980)$''-particle is expected to have a mass somewhat above 1~GeV.}, as well as contributions 
from heavier and/or more complicated multiparticle states which we have simply indicated by ``$\ldots$''. 
Our conventions are such that in eq.~(\ref{D55neu}) the single-particle meson states $\alpha$ 
($\alpha = \pi, \sigma^0_n, \dots$) are introduced with the Lorentz covariant normalization 
($b,b'=1,2,3$ are SU(2) isospin labels)
\beq 
\langle\alpha^b(\vec p)|\alpha^{b'}(\vec p\,')\rangle \, = \,
2E_\alpha(\vec p) \, (2\pi)^3 \delta_{b,b'}\delta(\vec p-\vec p\,')\, ,
\quad E_\alpha(\vec p) = \sqrt{\vec p\,^2+m_\alpha^2} \, .
\label{NORM}
\eeq 
The states $| {\cal P}_{\pi\pi}^{(I_3=0)} \pi \pi, E \rangle$ are two-pion states 
with total energy $E$, zero total three-momentum and zero total isospin
third component. In particular the symbol ${\cal P}_{\pi\pi}^{(I_3=0)}$ 
denotes the isospin projector onto the two-pion states with vanishing total third component ($I_3=0$). 
Here we recall that two possible states contribute which one has to sum over, namely 
the state with two neutral pions and that with one negatively and one positively charged pion.

Similarly the expression of the (infinite volume limit) spectral 
decomposition of $\Delta^\pm_{55}$ (eq.~(\ref{D55})) will read   
\beqn
&&\hspace{-0.9cm} \Delta^\pm_{55}= - \frac{\xi_\pi}{m^2_\pi} 
\langle\pi^\pm(\vec 0)|{\cal L}_5^{\rm Mtm}|\pi^3(\vec 0)\pi^\pm(\vec 0)\rangle
+ \label{D55cha} \\
&&- \frac{4\pi}{16\pi^3} \int_{2 m_\pi}^{4m_\pi} dE \frac{ k(E) }{E^2 - m^2_\pi}
  | \langle\pi^\pm(\vec 0)|{\cal L}_5^{\rm Mtm} | {\cal P}_{\pi\pi}^{(\pm,3)} 
    \pi \pi, E \rangle |^2 + R_{55}^{\pm}\nonumber \\
&&\hspace{-0.9cm}  R_{55}^{\pm}=- \sum_{n} \frac{ |\langle\pi^\pm(\vec 0)|{\cal L}_5^{\rm Mtm}|
        \sigma^\pm_n(\vec 0)\rangle|^2 }{m_{\sigma^\pm_n}^2 - m^2_\pi} +\ldots\, ,\label{REMPM}
\eeqn
with  ${\cal P}_{\pi\pi}^{(\pm,3)}$ the isospin projector onto the state 
of two pions having third component $\pm$ and 3. 

On general grounds one expects the numerically dominating contributions to 
$\Delta^{3,\pm}_{55}$ to come from the interaction of the propagating pion (the
initial and final particle in eq.~(\ref{D55})) with a neutral pion created from the vacuum,
i.e.\ from first term in the r.h.s.\ of eqs.~(\ref{D55neu}) and~(\ref{D55cha}). In the next 
sections we will show that this is indeed the case. The important observation is that 
in Mtm-LQCD with optimal critical mass these dominant tadpole contributions are O($m_\pi^2$)
and moreover equal for $\Delta^{3}_{55}$ and $\Delta^{\pm}_{55}$ up to (very small) 
O($m_\pi^4$) terms, hence they get canceled in the pion squared mass difference. 
The estimated leftover correction coming from the elastic two-pion cut as well as the 
remainders (eqs.~(\ref{REMNEU}) and~(\ref{REMPM})) will be shown to be tiny. 

We now examine in turn the various terms contributing to $\Delta^{3}_{55}$ and $\Delta^{\pm}_{55}$ 
and discuss their parametric behaviour in the small-$m_\pi^2$ regime. We shall see in particular that
for symmetry reasons the contributions from the various intermediate states to $\Delta^{\pm}_{55}$ tend 
to be negligible compared to those contributing to $\Delta^{3}_{55}$, with the exception of the tadpole 
term and the one coming from the lowest-lying two-pion intermediate state.

\subsubsection{Tadpole}
\label{sec:TAD}

That the first terms in eqs.~(\ref{D55neu}) and~(\ref{D55cha}) are equal to leading order 
in the chiral expansion follows, e.g.\ by reducing the neutral pion in the ket state, 
yielding~\footnote{In this Appendix with the symbol $\sim$ we denote equality up to O($a^2$) corrections, 
with $\stackrel{\rm SPT}{=}$ equality up to O($m_\pi^2$) terms 
and with $\stackrel{\rm SPT}{\sim}$ equality up to O($a^2$) and O($m_\pi^2$) corrections.}
\beq
\langle\pi^3(\vec 0)|{\cal L}_5^{\rm Mtm}|\pi^3(\vec 0)\pi^3(\vec 0)\rangle\stackrel{\rm SPT}{=}
\langle\pi^\pm(\vec 0)|{\cal L}_5^{\rm Mtm}|\pi^3(\vec 0)\pi^\pm(\vec 0)\rangle\, .\label{SPTRELL3}
\eeq
An explicit estimate of the tadpole terms can be given on the basis of the SPT relations 
\beq
\langle\pi^3(\vec 0)|{\cal L}_5^{\rm Mtm}|\pi^3(\vec 0)\pi^3(\vec 0) \rangle 
\stackrel{\rm SPT}{=} 
-\frac{1}{f_\pi^2} \langle \Omega |{\cal L}_5^{\rm Mtm}|\pi^3(\vec 0) \rangle =
-\frac{\xi_\pi}{f_\pi^2}\, ,\label{SPTREL2} \eeq
from which one gets
\beqn
&&\Delta^3_{55}\Big{|}_{tad}=
-\frac{\xi_\pi}{m^2_\pi} \langle\pi^3(\vec 0)|{\cal L}_5^{\rm Mtm}|\pi^3(\vec 0)\pi^3(\vec 0)\rangle\stackrel{\rm SPT}{=}
\frac{\xi_\pi^2}{f_\pi^2 m_\pi^2}\, ,\nn\\
&&\Delta^\pm_{55}\Big{|}_{tad}=
-\frac{\xi_\pi}{m^2_\pi} \langle\pi^\pm(\vec 0)|{\cal L}_5^{\rm Mtm}|\pi^3(\vec 0)\pi^\pm(\vec 0)\rangle
\stackrel{\rm SPT}{=} \frac{\xi_\pi^2}{f_\pi^2 m_\pi^2}\, .\label{TADES}
\eeqn
We remark that when the untwisted mass, $m_0$, is fixed 
at its optimal value, eq.~(\ref{OCM}) holds~\cite{FMPR}, implying 
\beq
\frac{\xi_\pi^2}{f_\pi^2 m_\pi^2}={\rm{O}}(m_\pi^2)\, , \label{OMPS}
\eeq
while in the opposite case one finds that this quantity tends to increase as we go towards the 
chiral limit, leaving large O($a^2/m_\pi^2$) cutoff effects in the neutral as well as in the 
charged lattice pion mass~\footnote{This result 
is in agreement with the finding of theoretical studies~\cite{AOBA,SHWU,FMPR} and indirectly 
confirmed by the observation of large, positive O($a^2$) lattice artifacts in the quenched Mtm-LQCD 
computations~\cite{CAN05,XLF05} of the charged pion mass and the corresponding decay constant 
(computed from the WTI relation $f_{\pi^{\pm}}=2\mu_q G_\pi/ m_{\pi^{\pm}}^{2}$).}. 

\subsubsection{Two-pion cut below the inelastic threshold}
\label{sec:TPCBIT}

In a finite three-dimensional volume, $V=L^3$, the integral over the two-pion cut can be 
rewritten as a sum over discrete states through the formal replacement 
\beq
\int_{2 m_\pi}^{4m_\pi} dE \frac{k(E)}{E^2 - m^2_\pi}\ldots \to  
\sum_{\ell=0}^{\ell_{max}}\frac{1}{\rho_V(E_\ell)} \frac{k(E_\ell)}
{E_\ell^2 - m^2_\pi}\ldots \, ,\label{REPL}
\eeq
where  $\ell_{\rm max}$ is the number of allowed two-pion levels that can be hosted in the volume $V$ 
below the inelastic threshold ($4m_\pi$). 
The relation between the allowed values of $\ell$ and the level energies 
is established combining eq.~(\ref{KAP}) with the formulae~\cite{LUSC,LUSCrhopipi} 
\beq
\ell\pi-\delta(k)=\phi(q) \, ,\qquad q=\frac{kL}{2\pi} \, ,\quad \ell \geq 0 \, ,\label{ENLEV}
\eeq
where $\delta(k)$ denotes the $s$-wave $\pi\pi$ phase-shift in the appropriate isospin  
channel and $\phi(q)$ is a known~\cite{LUSCrhopipi} kinematical function. 
Finally (see e.g.\ ref.~\cite{LMST2001}) 
\beq
\rho_V(E) = \frac{q \phi'(q) + k \delta'(k)}{4\pi k^2} \, E  \, , 
\label{RHOV}
\eeq
is the two-particle state density with $\delta'(k) = d\delta (k)/dk$ and $\phi'(q) = d\phi/dq$. 
At the volumes ($L = 2.1$ and $L=2.8$~fm) that will be considered in sect.~\ref{sec:NUMERICO} 
and for the lightest pion mass ($m_{\pi^\pm} \sim 300$~MeV) one has  $l_{\rm max}=2$.

$\bullet$ $\ell=0$ - The contribution of the $\ell=0$ term in the sum~(\ref{REPL}) can be computed 
in an almost model-independent way observing that the lowest energy level corresponds to have the two 
pions at rest with $E_0 = 2 m_\pi$~\footnote{Actually there is a finite size correction proportional to the 
($J=0$, $I=0$) $\pi\pi$ scattering length~\cite{LUSC}. To be precise one gets 
$E_0=2m_\pi-4\pi a_0^0 \,(m_\pi V)^{-1}(1+{\rm O}(1/L))$.}.   
Making use of the SPT's~(\ref{SPTRELL3})--(\ref{SPTREL2}) and taking into account the presence of the 
isospin projectors ${\cal P}_{\pi\pi}^{(I_3=0)}$ and ${\cal P}_{\pi\pi}^{(\pm,3)}$ in eqs.~(\ref{D55neu}) 
and~(\ref{D55cha}), respectively, one finds 
\beqn
&&\Delta^3_{55}\Big{|}_{\ell=0}=
- \frac{3 h_0^2 \xi_\pi^2}{(E_0^2 - m_\pi^2) f_\pi^4}\Big{|}_{E_0=2m_\pi}\, ,\label{LUZNEU}\\
&&\Delta^\pm_{55}\Big{|}_{\ell=0}= 
- \frac{2 h_0^2 \xi_\pi^2}{(E_0^2 - m_\pi^2) f_\pi^4}\Big{|}_{E_0=2m_\pi}\, ,\label{LUZPM}
\eeqn
where
\beq
h_0^2 = \frac{k(E_0)}{8\pi^2 \rho_V(E_0)}\Big{|}_{E_0=2m_\pi}= 
\frac{1}{2m_\pi V} \Big{(} 1 + {\rm O}(1/L) \Big{)} \, \label{H0_DEF}
\eeq
will be in the following approximated by its infinite volume limiting value, $(2m_\pi V)^{-1}$, which 
we remark is independent of the expression of the phase shift $\delta(k)$ appearing in $\rho_V(E)$.

Two observations are in order here. The first is that the quantities~(\ref{LUZNEU}) 
and~(\ref{LUZPM}) only differ because different isospin combinations of two-pion 
intermediate states contribute. The second is that, as in the tadpole case discussed above, 
both contributions are of O($m_\pi^2$), if eq.~(\ref{OMPS}) holds, i.e.\ if the critical
mass has been set to its optimal value. If this is not so, large and unequal, O($a^2/m_\pi^2$) 
cutoff effects will be left out in the neutral as well as in the charged lattice pion mass. 

$\bullet$ $\ell\geq 1$ - As we said, given the magnitude of $L$ we consider, in the energy interval 
$2 m_\pi<E<4m_\pi$ two-pion states with $\ell=1$ and $\ell=2$ will contribute with terms of the form
\beq
\Delta^b_{55}\Big{|}_{\ell} = - \frac{ 2h_\ell^2 }{ E_\ell^2 - m_\pi^2 } 
\Big{|} \langle\pi^b(\vec 0)| {\cal L}_5^{\rm Mtm} | {\cal P}_{\pi\pi}^{(...)} 
\pi\pi , E_\ell \rangle\Big{|}^2 \, ,
\quad 
h_\ell^2 = \frac{k(E_\ell)}{8\pi^2 \rho_V(E_\ell)}
\label{D55-ELL12}
\eeq
with $b$ equal to $3$ or $\pm$ and ${\cal P}_{\pi\pi}^{(...)}$ standing for ${\cal P}_{\pi\pi}^{(I_3=0)}$
or ${\cal P}_{\pi\pi}^{(\pm,3)}$, respectively. Evaluation of $\Delta^{3}_{55}|_{\ell=1,2}$ is possible 
only at the price of making some reasonable assumption about the energy behaviour of the pion matrix 
elements in eq.~(\ref{D55-ELL12}). The simplest thing is to assume that these matrix elements are 
constant in energy in the range $2 m_\pi<E<4m_\pi$. The quantities $h_\ell^2$ in eq.~(\ref{D55-ELL12}) 
can be straightforwardly evaluated in terms of $k(E_\ell)$ and $\rho_V(E_\ell)$ for the $\ell$-values of interest, 
once a reasonable parameterization of the phase-shift $\delta(k)$ (see sect.~\ref{sec:NUMERICO})
is inserted in eq.~(\ref{RHOV}).

Concerning $\Delta^{\pm}_{55}|_{\ell=1,2}$, the important remark is that these quantities
in the chiral regime are suppressed by a factor of $m_\pi^4$ with respect to their neutral counterparts
$\Delta^{3}_{55}|_{\ell=1,2}$. The result follows from eq.~(\ref{D55-ELL12}) by noting the
SPT relation (actually valid for all $\ell > 0$) 
\beq
\hspace{-0.01cm}i f_\pi \langle\pi^\pm(\vec 0)|{\cal L}_5^{\rm Mtm}| {\cal P}_{\pi\pi}^{(\pm,3)} \pi\pi ,\! E_\ell \rangle \!=\! 
\langle\Omega|[Q_A^\pm,{\cal L}_5^{\rm Mtm}]| {\cal P}_{\pi\pi}^{(\pm,3)} \pi\pi , \!E_\ell \rangle\! = \!
{\rm O}(m_\pi^2) \, .
\!\!\!\label{SPTDPM-ELL12}
\eeq
As a consequence, assuming the validity of this SPT for $m_\pi \simeq 300 \div 400$~MeV,
we shall neglect $\Delta^{\pm}_{55}|_{\ell=1,2}$ with respect to $\Delta^{3}_{55}|_{\ell=1,2}$
in our subsequent numerical estimates. 

We also observe that, if ${\cal L}_5^{\rm Mtm}$ itself is an operator of O($\mu_q$) = O($m_\pi^2$),
as it happens in the LO-$\chi$PT description of Mtm-LQCD with optimal critical 
mass~\footnote{In LO-$\chi$PT there is only one operator (see e.g.\ ref.~\cite{SHWU}) with the same 
quantum numbers as ${\cal L}_5^{\rm Mtm}$ (eq.~(\ref{L5Mtm})). Hence the condition
$\xi_\pi = \langle \Omega | {\cal L}_5^{\rm Mtm} | \pi^3(\vec 0) \rangle = {\rm O}(\mu_q)$
implies that the unique effective operator representing ${\cal L}_5^{\rm Mtm}$ at LO of $\chi$PT must 
itself be proportional to the quark mass. Obviously for matrix elements involving states with 
increasing rest-frame energy $E$ the LO-$\chi$PT description progressively looses its validity.}, 
the r.h.s.\ of eq.~(\ref{SPTDPM-ELL12}) is actually an O($m_\pi^4$) quantity for states with $E_\ell = {\rm O}(m_\pi)$, 
while $ \langle\pi^3(\vec 0)|{\cal L}_5^{\rm Mtm}| {\cal P}_{\pi\pi}^{(I_3=0)} \pi\pi , E_\ell \rangle$
is O($m_\pi^2$). In this situation one thus finds $\Delta^{\pm}_{55}|_{\ell=1,2}={\rm O}(m_\pi^6)$ 
and $\Delta^{3}_{55}|_{\ell=1,2}={\rm O}(m_\pi^2)$. The resulting $m_\pi^4$ suppression factor 
is in agreement with the general $m_\pi^2$-behaviour (implied by the SPT~(\ref{SPTDPM-ELL12})) 
of the charged pion contribution with respect to the neutral one. 

\subsubsection{The remainder}
\label{sec:TREM}

The remainder comprises the two-pion cut integral from $4m_\pi$ to $\infty$ and the 
contribution from all other possible single and multiparticle intermediate states. 

The first important observation is that all the contributions to $R^{\pm}_{55}$ are of 
O($m_\pi^4$). This result can be derived in close analogy to eq.~(\ref{SPTDPM-ELL12}) above,
namely by reducing the charged pion in the matrix elements (see eq.~(\ref{REMPM}))  
$\langle\pi^\pm(\vec 0)|{\cal L}_5^{\rm Mtm}| \alpha^\pm \rangle$, where $|\alpha^\pm\rangle$ 
is a state with a non-zero energy in the chiral limit (e.g.\ the one-particle scalar
state $\sigma_n^\pm$ appearing in eq.~(\ref{REMPM})). One gets, in fact  
\beq
i f_\pi \langle\pi^\pm(\vec 0)|{\cal L}_5^{\rm Mtm}| \alpha^\pm \rangle=
\langle\Omega|[Q_A^\pm,{\cal L}_5^{\rm Mtm}]| \alpha^\pm \rangle={\rm O}(m_\pi^2)\, .
\label{SPTDPM}
\eeq
A similar result does not hold for $R^{3}_{55}$ because the commutator 
$[Q_A^3,{\cal L}_5^{\rm Mtm}]$ does not vanish in the chiral limit. Based on
eq.~(\ref{SPTDPM}) we shall neglect the term $R^{\pm}_{55}$ with respect
to $R^{3}_{55}$ in the numerical estimates below.

In this context it should also be noted that the sum $\Delta^{\pm}_{55}|_{\ell=1} +
\Delta^{\pm}_{55}|_{\ell=2}  + R^{\pm}_{55}$, being
a negative quantity, yields a (tiny) positive contribution to the 
phenomenologically negative~\cite{LET,PROC} squared pion mass splitting. 
This term (see eqs.~(\ref{DMPPM})--(\ref{DMP3})), if included in $\Delta^{\pm}_{55}$, 
would go in the direction of increasing the negative contribution of $\zeta_\pi$ that 
is necessary in order to reproduce the observed value of $m_{\pi^3}^2|_{L}-m_{\pi^\pm}^2|_{L}$. 

Coming back to the evaluation of the remainder terms, we shall
hence limit below our discussion to $R^{3}_{55}$. The latter, we recall,
includes, besides terms from one-particle scalar intermediate
states explicitly shown in eq.~(\ref{REMNEU}), also a cut contribution of the form
\[
- \frac{4\pi}{16\pi^3} \int_{4 m_\pi}^{\infty} dE \frac{k(E) }{E^2 - m^2_\pi}
| \langle\pi^3(\vec 0)|{\cal L}_5^{\rm Mtm} | {\cal P}_{\pi\pi}^{(I_3=0)} 
  \pi \pi, E \rangle |^2 \, , 
\]
as well as negligible contributions from more complicated (and heavier) 
multiparticle intermediate states. The cut contribution above can be estimated by 
assuming the standard K\"allen--Lehmann (large energy) $1/E^2$ 
behaviour for the modulus square of the ${\cal L}_5^{\rm Mtm}$ matrix elements 
and by setting the coefficient factor in front to a value that matches
the contribution from the $\ell = 2$ two-pion intermediate state. 
As for the one-particle scalar state contribution, we make the plausible
assumption that it is of the same (very small, see sect.~\ref{sec:NUMERICO}) 
size as the cut contribution, and simply double the latter in order to get 
our numerical estimate of $R^{3}_{55}$.

Terms from multiparticle intermediate states of increasing mass are expected to give 
tiny and progressively smaller and smaller contributions, owing to explicit energy
denominators and the asymptotic high-energy behaviour of the matrix elements.

\subsection{Numerical considerations}
\label{sec:NUMERICO} 

In this subsection we want to provide a numerical estimate of the order of the magnitude of the 
difference $\Delta^{3}_{55}-\Delta^{\pm}_{55}$ relying on the results of the previous 
subsection and the numerical evaluation of $\xi_\pi$ given in sect.~\ref{sec:PROP}. 

We shall see that $\Delta^{\pm}_{55}$, which according to the considerations of sect.~\ref{sec:C1} 
is an O($m_\pi^2$) quantity, takes a value much smaller than $a^{-2} \Delta m_\pi^2 |_L^{\rm Mtm}$. 
Numerically $\Delta^{3}_{55}$ is of similar size as $\Delta^{\pm}_{55}$, as the parametrically O($m_\pi^0$) 
contributions from $R^{3}_{55}$ are tiny and actually smaller than the O($m_\pi^2$) terms.

These facts imply (see eqs.~(\ref{DMPPM}) and~(\ref{DMP3})) that lattice corrections to 
the squared charged pion mass are also small and of order $a^2m_\pi^2$ (or $a^4$), while the main 
contribution to the lattice artifacts of the squared neutral pion mass comes from $\zeta_\pi$, 
as announced in the text (sect.~\ref{sec:PMMTM}). As we already said, in the difference
$\Delta^{3}_{55} - \Delta^{\pm}_{55}$ the contribution coming from the tadpole terms exactly cancels. 

Using the value~(\ref{XIPI_DETER}) of $\xi_\pi$ and the estimate 
$h_0^2=(2m_\pi V)^{-1}$, we can provide an explicit numerical evaluation of the three 
contributions to $\Delta^{3}_{55}$ and $\Delta^{\pm}_{55}$ described in sects.~\ref{sec:TAD}, 
\ref{sec:TPCBIT} and~\ref{sec:TREM}. To arrive at this result we also need to know 
the energies of the few lowest $\pi\pi$ levels living in our finite simulation box at the  
actual value of the simulated pion masses. In the situation corresponding to the lattice 
data used for our best estimate of $\xi_\pi$, we have $m_\pi\simeq 300$~MeV 
and $L \sim 2.1$~fm. We checked that results do not change significantly with the volume
by repeating the same analysis for $L \sim 2.8$~fm. In order to compute $k_\ell$ and hence 
$E_\ell = 2 \sqrt{k_\ell^2 + m_\pi^2}$ from eq.~(\ref{ENLEV}), as well as $\rho_V(E_\ell)$ 
from eq.~(\ref{RHOV}), we parameterize the $s$-wave $\pi\pi$ phase shift in the form
(suggested by LO-$\chi$PT~\cite{GASSMEIS}) 
\beq
\delta(k) = \frac{2k}{E}\cdot \frac{2 E^2 - m_\pi^2}{32\pi f_0^2}
\, , \quad \frac{E^2}{4} = k^2 + m_\pi^2 \, ,
\label{PHSH}
\eeq
with $f_0 = 86$~MeV being the chiral limit value of the pion decay constant
(as obtained from recent ETMC studies~\cite{PROC,LET}). The formula~(\ref{PHSH}) 
(with $m_\pi$ at its physical value $140$~MeV) describes rather well the experimental 
$\pi\pi$-scattering data for center-of-mass energy $E$ in the range $2m_\pi$ to $4m_\pi$ 
(see e.g.\ the phase shift data compilation in ref.~\cite{GASSMEIS}).

Following the discussion in sects.~\ref{sec:TAD},  \ref{sec:TPCBIT} and~\ref{sec:TREM}, in terms of 
the reference scale $a_{\beta=3.9}^{-1}$, whose value is currently estimated~\cite{PROC} to be about 
2.3~GeV, we obtain the following numerical results.

$\bullet$ - {\it Tadpole}
\beq
\Delta^3_{55}\Big{|}_{tad}=\Delta^\pm_{55}\Big{|}_{tad}
\stackrel{\rm SPT}{=} \frac{\xi_\pi^2}{f_\pi^2 m_\pi^2} \sim  0.00050^{+0.00032}_{-0.00024} \, a_{\beta=3.9}^{-4} 
\, , \label{TADESN}
\eeq
where the quoted error reflects the uncertainty on $\xi_\pi$ from eq.~(\ref{XIPI_DETER}).

$\bullet$ - {\it Two-pion cut below the inelastic threshold}. For the $\ell=0$ term one obtains  
\beqn
&&\Delta^3_{55}\Big{|}_{\ell=0}=
- \frac{3 h_0^2 \xi_\pi^2}{3m_\pi^2 f_\pi^4}\sim - ( 0.000063^{+0.000040}_{-0.000030} ) \, a_{\beta=3.9}^{-4} 
\, ,\label{LUZNEUN}\\
&&\Delta^\pm_{55}\Big{|}_{\ell=0}=
-\frac{2 h_0^2 \xi_\pi^2}{3m_\pi^2 f_\pi^4}\sim-(0.000042^{+0.000027}_{-0.000020})\, a_{\beta=3.9}^{-4}
\, .\label{LUZPMN}
\eeqn
For the $\ell>0$ levels one gets smaller and smaller contributions. A reasonable estimate 
of their order of magnitude is $-0.000020 a_{\beta=3.9}^{-4}$ and 
$-0.000015 a_{\beta=3.9}^{-4}$ for $\Delta^3_{55}|_{\ell}$ at
$\ell=1$ and $\ell=2$, respectively, to which we attach 
a generous 50--60\% relative uncertainty. 
We neglect here $\Delta^\pm_{55}|_{\ell=1,2}$, which according to SPT arguments, are 
significantly smaller ($m_\pi^4$ suppressed) than their neutral counterparts. 

$\bullet$ - {\it Remainder}. According to the results of subsect.~\ref{sec:TREM}, we can limit 
consideration to $R^3_{55}$. Despite the difficulty of evaluating all the terms contributing to $R^3_{55}$, 
following the procedure discussed in sect.~\ref{sec:TREM}, we can reasonably 
estimate it to be about $-0.000080 a_{\beta=3.9}^{-4}$, 
with, as before, a 50--60\% relative statistical uncertainty. We thus find that $R^3_{55}$ is 
of the same order of magnitude as $\Delta^3_{55}|_{\ell=0}$ or $\sum_{\ell=1}^{2} \Delta^3_{55}|_{\ell}$. 

$\bullet$ - {\it Total}. Putting everything together we finally get 
\beqn
&&\Delta^3_{55}\simeq +(0.00032 \pm 0.00019 \pm 0.00008) a_{\beta=3.9}^{-4} \, ,\label{D55T}\\
&&\Delta^{\pm}_{55} \simeq +(0.00046 \pm 0.00027 \pm 0.00005) a_{\beta=3.9}^{-4} \, ,\label{DPMT}
\eeqn
and
\beq
\Delta^3_{55}-\Delta^\pm_{55} \simeq -(0.00014 \pm 0.00008 \pm 0.00004) a_{\beta=3.9}^{-4} \, .\label{DDT}
\eeq
The two ``errors'' quoted in the equations above reflect the uncertainty (dominantly of statistical nature) 
on the value of $\xi_\pi$ from sect.~\ref{sec:PROP} and the systematic indetermination inherent to the 
assumptions we made above about the energy behaviour of the matrix elements of 
${\cal L}_5^{\rm Mtm}$, respectively.

For the sake of comparison here we only recall that (as one can infer from Fig.~\ref{fig:figMPISPLIT}
and the value $r_0/a|_{\beta=3.9} \simeq 5.2$) at $m_\pi \simeq 300$~MeV direct lattice measurements 
yield $a_{\beta=3.9}^2 \Delta m_\pi^2|_{L;\beta=3.9}^{\rm Mtm} \simeq -0.0067(20)$,
i.e.\ a number much bigger than $a_{\beta=3.9}^{4} \Delta^{3}_{55}$, or $a_{\beta=3.9}^{4} \Delta^{\pm}_{55}$. 
Comparisons of this kind are performed and discussed more thoroughly in sect.~\ref{sec:MPISPLIT}. 

\subsection{Estimating $\xi_\pi$ from lattice two-point correlators}   
\label{sec:PROP}

We describe in this section the analysis, based on the Symanzik expansion, by which we arrive at a  
numerical estimate of the crucial continuum quantity $\xi_\pi$ (eqs.~(\ref{DEFXI})--(\ref{OCM})) 
that has been employed in sect.~\ref{sec:C1} to parametrize the various terms contributing to 
$\Delta^{\pm}_{55}$ and $\Delta^{3}_{55}$. Since in this numerical analysis we will be 
exploiting $N_f=2$ Mtm-LQCD data at sufficiently small $m_\pi \simeq m_{\pi^\pm}|_L$ around 300~MeV, 
we feel safe to use SPT's to relate among themselves some of the (continuum QCD)
matrix elements that appear in the Symanzik description of lattice quantities. The SPT approximation
is correct up to O($m_\pi^2$) relative corrections, whose relevance is checked by going 
to heavier quark masses up to $m_{\pi^\pm}|_L \leq 450$~MeV. By extending the analysis from 
data at lattice resolution $a^{-1} = 2.3$~GeV ($\beta=3.90$) to those at $a^{-1} = 2.9$~GeV 
($\beta=4.05$) and comparing (at the lowest-lying pion mass) results at box linear size $L \sim 2.1$~fm 
with those at $L \sim 2.8$~fm, we have checked that within our (large) statistical errors the numerical 
estimates relevant for this paper display no significant discretization and/or finite size effects. 	 

The key correlator we consider is~\footnote{We recall that all the correlators we shall be dealing 
with are ``connected'' in the strict sense of Quantum Field Theory, though some contribution 
only through gluon exchanges.}
\beq
C_{PS}^{L}(t)= \sum_{\vec x} \frac{1}{2} \langle P^3(\vec x, t) 
S^0(0)\rangle\Big{|}_L-L^3 \frac{1}{2} \langle P^3(0)\rangle\langle S^0(0)\rangle\Big{|}_L \, ,
\qquad t>0 \, ,
\eeq
which in the continuum limit is an O($a$) quantity because it violates parity (and isospin) invariance.
The Symanzik expansion of its renormalized counterpart reads 
\beqn
&&Z_{S^0} Z_{P}\, C_{PS}^{L}(t)
=a\langle\int d{\vec x} \hat P^3(\vec x, t)\int d^4 y \,{\cal L}_5^{\rm Mtm}(y) \frac{1}{2}\hat S^0(0)\rangle
+{\mbox O}(a^3) =\nn\\
&&= a\hat G_\pi\frac{\xi_\pi}{m_\pi^2} \langle\pi^3|\frac{1}{2}\hat S^0|\pi^3\rangle\frac{e^{-m_\pi t}}{2m_\pi}  +\nn\\
&&+ a\hat G_\pi  \langle\pi^3|{\cal L}_5^{\rm Mtm}|\sigma_{\rm eff}\rangle
\langle\sigma_{\rm eff}|\frac{1}{2}\hat S^0|\Omega\rangle \frac{1}{m_{\sigma_{\rm eff}}^2-m_\pi^2}
\Big{[}\frac{e^{-m_\pi t}}{2m_\pi}-\frac{e^{-m_{\sigma_{\rm eff}} t}}{2m_{\sigma_{\rm eff}}}\Big{]} +\nn\\
&&+ a\frac{\xi_\pi}{m_\pi^2} \langle\pi^3|\hat P^3|\sigma_{\rm eff}\rangle
\langle\sigma_{\rm eff}|\frac{1}{2}\hat S^0|\Omega\rangle
\frac{ e^{-m_{\sigma_{\rm eff}} t} }{ 2m_{\sigma_{\rm eff}} } +{\mbox O}(a^3) \, ,
\label{CPSYM1}\eeqn
where $\hat{P}^3$ and $\hat S^0$ denote the renormalized operators ${P}^3$ and $S^0$, 
respectively, and $G_\pi$ was defined in eq.~(\ref{G-SIGMA-DEF}). 
The symbol $|\sigma_{\rm eff}\rangle$ represents the lightest (isosinglet) 
scalar state that can be created from the vacuum by the operator $S^0$. For the quark mass 
values and in the $t$-range of interest for the present analysis of $C_{PS}^{L}(t)$ 
(because of the unfavourable signal-to-noise ratio inherent in the evaluation of the 
quark disconnected piece of the correlator), we are in fact sensitive within errors only to the 
neutral pion and $|\sigma_{\rm eff}\rangle$ intermediate states. In a finite box and for 
sufficiently small quark masses, the symbol $|\sigma_{\rm eff}\rangle \langle\sigma_{\rm eff}|$ 
in the r.h.s.\ of eq.~(\ref{CPSYM1}) is to be interpreted as 
\beqn
|\sigma_{\rm eff}\rangle \langle\sigma_{\rm eff}| = h_0^2 \sum_{b=1}^3 | \pi^b\pi^b, m_{\sigma_{\rm eff}}\rangle
\langle \pi^b\pi^b, m_{\sigma_{\rm eff}}|  \, , \quad m_{\sigma_{\rm eff}}\simeq 2m_\pi \, .
\label{SIGMAEFF}
\eeqn
In the following we shall write simply $|\sigma\rangle$ for $|\sigma_{\rm eff}\rangle$.

The three terms in the r.h.s.\ of eq.~(\ref{CPSYM1}) come from the three different possible 
time-orderings of the inserted operators. We have only retained the pole contributions from 
the two lightest states and the associated partially disconnected terms that must go along 
with them in order to comply with Lorentz invariance. Since in Mtm-LQCD with optimal 
critical mass all the leading terms (linear in $a$) in eq.~(\ref{CPSYM1}) are of the same order in $m_\pi$,
we have ignored here the O($a$) lattice corrections to the operators $S^0$ and 
$P^3$, because they would give contributions to the Symanzik expansion 
suppressed by an extra (relative) factor $m_\pi^2$~\footnote{This is seen by noting that 
$\Delta_1 S^0 \propto a\mu_q P^3$, while $\Delta_1 P^3$ comprises a term proportional 
to $a\mu_q S^0$ and another one proportional to $F_{\mu\nu} F_{\mu\nu}$. 
This last term has O($\mu_q$) matrix elements between states involving only the vacuum and
one or two neutral pions, as one can check by means of SPT's.} and are 
numerically small for the considered range of quark mass and $t$-values.

With the help of eqs.~(\ref{SPTREL2}) and~(\ref{SIGMAEFF}) as well as the SPT relations
\beqn
&& \langle\pi^3| \frac{1}{2} \hat S^0|\pi^3\rangle 
\stackrel{\rm SPT}{=} \frac{1}{f_\pi} \hat G_\pi \, , \label{SPTREL3}\\
&&\langle\pi^3| \hat P^3|\sigma\rangle \stackrel{\rm SPT}{=} 
-\frac{1}{f_\pi} \langle\Omega| \frac{1}{2} \hat S^0|\sigma\rangle
\, ,\label{SPTREL4}
\eeqn
the Symanzik expansion~(\ref{CPSYM1}) can be rewritten in the simpler form 
\beqn
\hspace{-.8cm}&&
Z_P Z_{S^0} C_{PS}^{L}(t) 
\stackrel{\rm SPT}{=} 
a \frac{e^{-m_\pi t}}{2m_\pi} \,
\frac{ \hat G_\pi^2 \xi_\pi }{ f_\pi m_\pi^2 } \Big[ 1 -
\frac{3h_0^2}{f_\pi^2} \frac{ m_\pi^2 }{ m_\sigma^2 - m_\pi^2 } \Big] + \nn \\
\hspace{-.8cm}&&
- a \frac{e^{-m_\sigma t}}{2m_\sigma} \, 
\frac{ \hat G_\pi^2 \xi_\pi }{ f_\pi m_\pi^2 } \frac{3h_0^2}{f_\pi^2} 
\Big[ 1 - \frac{ m_\pi^2 }{ m_\sigma^2 - m_\pi^2 } \Big]
\; + \; {\mbox O}(a^3) \, ,
\label{CPSYM2}\eeqn
where, as we shall see below, all the quantities, but $\xi_\pi$ and the negligible O($a^3$)
terms, can be directly evaluated from lattice data. Solving, in fact, the resulting equations 
yields an estimate of for $\xi_\pi$ for each given quark mass $\mu_q$ and lattice resolution $a$.

\subsubsection{Evaluating $\xi_\pi$}   
\label{sec:NUM1}

Several two-point Green functions in the neutral pseudoscalar (PS) channel (where also 
quark-disconnected diagrams contribute) have been evaluated, among which 
\beqn
&& C_{PP}^{L}(t) = a^3\sum_{\vec x} \langle P^3(x) P^3(0) \rangle\Big{|}_{L}
   - L^3 \langle P^3(0) \rangle\Big{|}_{L} \langle P^3(0) \rangle\Big{|}_{L}  \, ,
\label{LAT_PP}
\\
&& C_{SS}^{L}(t) = a^3\sum_{\vec x} \frac{1}{4} \langle S^0(x) S^0(0) \rangle\Big{|}_{L}
   - L^3 \frac{1}{4} \langle S^0(0) \rangle\Big{|}_{L} \langle S^0(0) \rangle\Big{|}_{L}  \,
\label{LAT_SS}
\eeqn
and
\beqn
C_{PS}^{L}(t) = a^3\sum_{\vec x} \frac{1}{2} \langle P^3(x) S^0(0) \rangle\Big{|}_{L}
   - L^3 \frac{1}{2} \langle P^3(0) \rangle\Big{|}_{L} \langle S^0(0) \rangle\Big{|}_{L}  \, ,
\label{LAT_PS}
\\
C_{SP}^{L}(t) = a^3\sum_{\vec x} \frac{1}{2} \langle S^0(x) P^3(0) \rangle\Big{|}_{L}
   - L^3 \frac{1}{2} \langle S^0(0) \rangle\Big{|}_{L} \langle P^3(0) \rangle\Big{|}_{L}  \, ,
\label{LAT_SP}
\eeqn
It should be noted that in connected correlators the mixing of the bare operators 
$P^3$ and $S^0$ with the identity does not contribute.

All lattice correlators are simultaneously fitted to simple two-state ansatz of the form
\beq
C_{PS}^{L}(t) = c_5 c_1 \frac{\exp(-m_{\pi^3}|_L t)}{2 m_{\pi^3}|_L} +
                       d_5 d_1 \frac{\exp(-m_\sigma|_L t)}{2 m_\sigma|_L} \, ,
\label{fit_C_PS}
\eeq 
\beq
C_{PP}^{L}(t) = c_5 c_5 \frac{\exp(-m_{\pi^3}|_L t)}{2 m_{\pi^3}|_L} +
                       d_5 d_5 \frac{\exp(-m_\sigma|_L t)}{2 m_\sigma|_L} \, ,
\label{fit_C_PP}
\eeq
\beq
C_{SS}^{L}(t) = c_1 c_1 \frac{\exp(-m_{\pi^3}|_L t)}{2 m_{\pi^3}|_L} +
                       d_1 d_1 \frac{\exp(-m_\sigma|_L t)}{2 m_\sigma|_L} \, .
\label{fit_C_SS}
\eeq
Comparing the fit ansatz for the correlators $C_{PS,SS,PP}^{L}$ with 
the corresponding Symanzik expansions written to the leading order in $a$ and in 
the form where only the neutral pion and/or the lightest scalar state contributions 
are retained (consistently with the numerical fact
that within statistical errors only the contribution of these two states is visible),
one obtains the following identifications 
\beq
c_5 \sim \langle \Omega | P^3 | \pi^3(\vec 0) \rangle \, , \qquad
d_1 \sim \langle \Omega | \frac{1}{2}S^0 | \sigma \rangle\, , \label{IDF_A}
\eeq 
as well as
\beq
Z_{S^0} Z_P c_1 c_5 
\stackrel{\rm SPT}{\sim}
a \frac{ \hat G_\pi^2 \xi_\pi }{ f_\pi m_\pi^2 } \Big[ 1 -
\frac{3h_0^2}{f_\pi^2} \frac{ m_\pi^2 }{ m_\sigma^2 - m_\pi^2 } \Big]\, .
\label{IDF_B}
\eeq
A similar equation between $Z_{S^0} Z_P d_1 d_5$ and the coefficient of the 
$e^{-m_\sigma t}/(2m_\sigma)$ term in the r.h.s.\ of eq.~(\ref{CPSYM2}) can be 
obtained. This relation is, however, of minor interest given the substantial statistical error 
we have on the fitted coefficient $d_5$. Finally the relevant dimensionless ratios $Z_P/Z_{S^0}$ 
and $3h_0^2/f_\pi^2$ entering eq.~(\ref{IDF_B}) can be estimated via the formulae 
\beq
\frac{Z_P}{Z_{S^0}} \sim |\frac{c_5}{G_{\pi^\pm}|_L}| \, , \qquad
\frac{h_0^2}{f_\pi^2} 
\stackrel{\rm SPT}{\sim} \frac{1}{3}
\Big{(}\frac{d_1}{G_{\pi^\pm}|_L} \Big{)}^2\, ,
\label{IDF_aux}
\eeq
where  
\beqn
&&G_{\pi^\pm}|_L = \langle \Omega | P^\pm | \pi^\pm (\vec 0) \rangle|_L \, ,\label{GPPM}\\	
&&P^\pm = (P^1 \pm i P^2)/\sqrt{2} \, .
\label{PPM}
\eeqn
The second of the eqs.~(\ref{IDF_aux}) is a consequence of the SPT~(\ref{SPTREL3}) and the relation
\beq
Z_P d_1 \stackrel{\rm SPT}{\sim} \sqrt{3} h_0 \langle \pi^3(\vec 0) | \frac{1}{2}\hat{S}^0 | \pi^3(\vec 0) \rangle
\label{DREL}
\eeq
which in turn follows from the identification of $\sigma$ as a two pion state (see eq.~(\ref{SIGMAEFF})). 
It should also be recalled that both the renormalized operators $\hat S_0$ and $\hat P^\pm$
are related to their bare (subtracted) counterparts through the renormalization constant $Z_P$, which 
would then cancel in the ratio $d_1/G_{\pi^\pm}|_L$ if we were to write it in terms of renormalized quantities. 

Combining together the relations~(\ref{IDF_A}), (\ref{IDF_B}) and~(\ref{IDF_aux}), one
arrives at an explicit formula for $\xi_\pi$, namely
\beq
a \xi_\pi 
\stackrel{\rm SPT}{\sim}
c_1  \frac{ f_{\pi^\pm} m_{\pi^\pm}^2 }{ G_{\pi^\pm} }  \Big{[} 1 - 
     \frac{ d_1^2 }{ 3G_{\pi^\pm}^2 } \frac{ 3m_{\pi^\pm}^2 }{ m_\sigma^2 - m_{\pi^\pm}^2 }  
     \Big{]}^{-1} \Big{|}_L \, ,
\label{XIPI_FORM}
\eeq
where the charged pion sector O($a^0$) quantities $m_{\pi^\pm}|_L$, $G_{\pi^\pm}|_L$ and
\beq
f_{\pi^\pm}\Big{|}_L \equiv 2 \mu_q (G_{\pi^\pm} m_{\pi^\pm}^{-2})\Big{|}_L  \label{fpiINDIR}
\eeq
can be evaluated much more accurately (at level of 1\% or better) than the corresponding 
neutral sector quantities, $m_\sigma|_L$, $d_1$ and $c_1$ (we recall that $c_1$ 
vanishes linearly in $a$ as $a \to 0$). 
{}From the analysis of the neutral pseudoscalar channel correlators specified in
eqs.~(\ref{LAT_PP}) to~(\ref{LAT_SP}), we can extract $m_\sigma|_L$ that, 
within statistical errors (ranging from 10\% to 20\% depending on statistics) 
turns out to be equal to $2m_{\pi^\pm}|_L$ at all the values of $\mu_q$ and 
at the two different lattice resolutions we consider (see Table~\ref{tabC} below). Using the 
relation~(\ref{fpiINDIR}) and setting $m_\sigma|_L = 2m_{\pi^\pm}|_L$, 
eq.~(\ref{XIPI_FORM}) takes the simple form 
\beq
a \xi_\pi 
\stackrel{\rm SPT}{\sim}
c_1 2\mu_q \left[ 1 
            - \frac{d_1^2}{3 G_{\pi^\pm}|_L^2} \right]^{-1} \, ,
\label{XIPI_FORM2}
\eeq
or alternatively
\beq
a \xi_\pi
\stackrel{\rm SPT}{\sim}
c_1 2\mu_q \left[ 1
            - \frac{h_0^2}{f_{\pi^\pm}|_L^2} \right]^{-1} \, .
\label{XIPI_FORM3}
\eeq
The latter expression of $\xi_\pi$ is exploited to obtain the numerical estimates 
presented below.

\subsubsection{Numerical results}   
\label{sec:NUM2}

We report in Table~\ref{tabC} in lattice units the results for the quantities entering
eqs.~(\ref{XIPI_FORM})--(\ref{XIPI_FORM3}). These numbers are obtained from the analysis of
the correlators~(\ref{LAT_PP})--(\ref{LAT_SP}), as well as their charged pseudoscalar meson analogs, 
evaluated on the gauge configurations taken from the $N_f=2$ ETMC ensembles with
the following bare parameters 
\beqn
\hspace{-0.8cm} \beta=3.9 \; (a \simeq 0.09~{\rm fm})\, , \; 
L/a=24 \, : & a\mu_q = 0.0040, 0.0064, 0.0085 \, ,
\label{b39L24}\\ 
\hspace{-0.8cm} \beta=3.9 \; (a \simeq 0.09~{\rm fm})\, , \; 
L/a=32 \, : & a\mu_q = 0.0040 \, ,
\label{b39L32}\\ 
\hspace{-0.8cm} \beta=4.05 \; (a \simeq 0.07~{\rm fm})\, , \; 
L/a=32 \, : & a\mu_q = 0.0030, 0.0060 \, .
\label{b405L32} 
\eeqn
The values of $a\mu_q$ above are known~\cite{PROC,ETMCLQM} to 
correspond to quark mass values (in the $\overline{MS}$ scheme at a scale
of 2~GeV) in the range from 20 to 40~MeV. At $\beta=3.9$ and $a\mu_q=0.0040$ data at two 
different volumes allow to check that our estimates of $\xi_\pi$ are not significantly affected 
by finite size effects. The results at the two $\beta$-values can be compared by expressing quantities in units 
of $r_0$ (extrapolated to zero quark mass limit) with the help of the estimates~\cite{PROC,ETMC_long})
\beq
r_0/a|_{\beta=3.9}=5.22(2) \, , \qquad
r_0/a|_{\beta=4.05}=6.61(3) \, . \label{R0_VAL}
\eeq
In order to ease such a comparison, for the case of $c_1$, which is a dimension two and O($a$) 
quantity, we also quote the value of $c_1 r_0^3 a^{-1}$, the latter being 
by construction a dimensionless and O($a^0$) quantity~\footnote{The statistical error 
on $c_1 r_0^3 a^{-1}$ is completely dominated by that on $a^2c_1$.}.
For the key quantity $\xi_\pi$ we quote for the same reasons $r_0^4 \xi_\pi$.
In Table~\ref{tabC} we also give for each ensemble the number of measurements  
(\# meas.) of the neutral pseudoscalar meson channel correlators. 

\begin{table}[htb]
  \centering
  \begin{tabular*}{1.\linewidth}{@{\extracolsep{\fill}}lcccc}
    \hline\hline
 $(\beta , a\mu_q)|_{L/a}$  &  
 $(3.9,.0040)|_{24}$  &  $(3.9,.0064)|_{24}$  & $(3.9,.0085)|_{24}$  &  $(3.9,.0040)|_{32}$  \\
    \hline\hline

\# meas.        &        880       &      240         &     248       &   184   \\
    \hline
$am_{\pi^\pm}|_L$ &      0.1359(7)   &     0.1694(4)    &    0.1940(5)  &  0.1338(3) \\
    \hline
$a^2G_{\pi^\pm}|_L\sqrt{2}$ &  0.1501(25)   &     0.1581(16)    &    0.1643(15)  &  0.1484(11) \\
    \hline
$am_{\pi^3}|_L$ &      0.103(4)   &     0.134(10)    &    0.163(8)  &   0.107(7) \\
    \hline
$am_{\sigma}|_L$ &     0.234(25)   &    0.391(46)    &    0.449(47)  &   0.284(47) \\
    \hline
$a^2 c_1 \sqrt{2}$        &     -0.021(4)   &    -0.018(6)    &    -0.011(6)   &   -0.019(5) \\
    \hline
$a^2 d_1 \sqrt{2}$        &     -0.12(2)   &    -0.22(5)    &    -0.25(6)   &   -0.15(3) \\
    \hline
$c_1 r_0^3 a^{-1} \sqrt{2}$  &     -3.0(6)     &    -2.6(9)     &   -1.6(9)     &   -2.7(7)   \\
    \hline
$\xi_\pi r_0^4 $  &     -0.10(2)     &    -0.14(5)     &   -0.10(4)     &   -0.08(2)   \\
    \hline\hline
 $(\beta , a\mu_q)|_{L/a}$  &  
                              $(4.05,.0030)|_{32}$  & $(4.05,.0060)|_{32}$  &   &   \\
    \hline
\# meas.       &         192         &     188     &   &    \\
    \hline
$am_{\pi^\pm}|_L$ &      0.1038(6)   &   0.1432(6) &   &    \\
    \hline
$a^2G_{\pi^\pm}|_L\sqrt{2}$ &   0.0898(18)   &   0.0972(12) &   &    \\
    \hline
$am_{\pi^3}|_L$ &      0.090(6)   &   0.125(6) &   &    \\
    \hline
$am_{\sigma}|_L$ &      0.231(37)   &   0.232(55) &   &    \\
    \hline
$a^2 c_1 \sqrt{2}      $ &      -0.014(3)    &   -0.010(6) &   &    \\
    \hline
$a^2 d_1 \sqrt{2}       $ &      -0.10(2)    &   -0.09(3) &   &    \\
    \hline
$c_1 r_0^3 a^{-1} \sqrt{2}$  &      -4.0(9)       &   -2.9(1.7)    &   &    \\
    \hline
$\xi_\pi r_0^4 $  &     -0.13(2)       &   -0.16(9)    &   &    \\
    \hline\hline
  \end{tabular*}
  \caption[tabC]{Quantities entering eqs.~(\ref{XIPI_FORM})--(\ref{XIPI_FORM3}).
The labels for the parameter sets have self-explanatory names and the corresponding columns
follow the same order as the list of parameter sets in eqs.~(\ref{b39L24})--(\ref{b405L32}). 
The statistical errors come from a standard bootstrap analysis on blocked data, so as
to properly take into account autocorrelation of consecutive measurements.}
 \label{tabC}
\end{table}

Inspection of Table~\ref{tabC} shows that the most accurate results are
obtained for $\beta=3.9$, $a\mu_q=0.0040$, where the number
of measurements of neutral pseudoscalar meson channel correlators is the largest 
and we also have data on two different physical volumes. From the results
at $\beta=3.9$, $a\mu_q=0.0040$ and using eq.~(\ref{XIPI_FORM3}),
we obtain our best $\xi_\pi$ estimate, namely
\beq
\xi_\pi 
\stackrel{\rm SPT}{\sim}
-0.00014(4) \, a_{\beta=3.9}^{-4}  \quad \Leftrightarrow \quad
r_0^4 \xi_\pi
\stackrel{\rm SPT}{\sim}
-0.10(3) \, ,
  \label{XIPI_DETER}
\eeq
from which eqs.~(\ref{TADESN})--(\ref{DDT}) 
follow. One checks in Table~\ref{tabC} that this result remains stable within statistical 
errors at increasing values of the quark mass.

If we were to use for $\xi_\pi$ the expression~(\ref{XIPI_FORM2}), instead
of eq.~(\ref{XIPI_FORM3}), results consistent with eq.~(\ref{XIPI_DETER})
would be obtained, though with a larger statistical errors due to the
fact that $d_1/(\sqrt{3}G_{\pi^\pm})|_L$ is a quantity of order one
with a sizeable uncertainty from $d_1$, the effect of which is enhanced in the factor 
$\Big{(}1 - \frac{d_1^2}{3 G_{\pi^\pm}|_L^2} \Big{)}^{-1}$. For instance, at $\beta=3.9$, 
$a\mu_q=0.0040$ and $L/a=24$ (the statistically most precise data set) 
we would get $\xi_\pi =  -0.00015(5) \, a_{\beta=3.9}^{-4}$. 

Results about the various quantities entering eqs.~(\ref{XIPI_FORM})--(\ref{XIPI_FORM3}) are
also shown in Table~\ref{tabC}, as a check of their quark mass dependence and scaling with $a$.
For instance, in the case of $c_1$ we find by inspection that the values of $c_1 r_0^3 a^{-1}$ 
display within errors good scaling and no visible quark mass dependence. This last property is 
consistent with the use of SPT's we made in order to arrive at eq.~(\ref{XIPI_FORM2}). The 
situation for $d_1$ is similar, but with somewhat increasing statistical uncertainties at higher 
quark masses. The statistically very precise matrix element $G_{\pi^\pm}|_L$, shows instead 
a mild quark mass dependence, as its value is seen to change by not more than 10\% in the quark 
mass range we have considered.


\vspace{.3cm}
{\bf{Acknowledgments - }} Illuminating discussions with M. Testa are gratefully
acknowledged. We are also indebted to M.~L\"uscher for stimulating remarks
and P.~Weisz for a technical discussion. 
We wish to thank the rest of ETMC for the interest in this 
work and for a most enjoyable and fruitful collaboration. This work has been supported 
in part by the DFG Sonder\-for\-schungs\-be\-reich/Transregio SFB/TR9-03, DFG project
JA 674/5-1 and the EU Integrated Infrastructure Initiative Hadron Physics (I3HP) 
under contract RII3-CT-2004-506078 and by the EU Contract MRTN-CT-2006-035482 
``FLAVIAnet''.  We also acknowledge the DEISA Consortium (co-funded by
EU under the FP6 project 508830) support within the DEISA Extreme
Computing Initiative (www.deisa.org).  G.C.R. and R.F. thank MIUR (Italy)
for partial financial support under the contracts PRIN04 and PRIN06.

\end{document}